\documentclass[10pt]{article}

\usepackage[dvips]{epsfig} % this package is for manipulating postscript graphics
\usepackage{amssymb,amsmath,amsthm,mathrsfs} %this package contains all the special ams fonts etc.
\usepackage{graphicx}

\usepackage[authoryear,sort&compress]{natbib}
\usepackage{url}
\usepackage{enumerate}
\usepackage{colordvi,color,pspicture}
\usepackage{psfrag}

\graphicspath{{Figures/}}

\newcommand{\Y}{\mathcal{Y}}
\newcommand{\X}{\mathcal{X}}
\newcommand{\mZ}{\mathcal{Z}}
\newcommand{\U}{\mathcal{U}}
\newcommand{\I}{\mathbf{I}}

\newcommand{\N}{\mathcal{N}}

\newcommand{\R}{\mathbb{R}}

\newcommand{\E}[1]{\mathbf{E}\,#1}
\newcommand{\Var}[1]{\mathbf{Var}\,#1}
\newcommand{\Cov}[1]{\mathbf{Cov}\,#1}

\newcommand{\ve}{\varepsilon}
\newcommand{\lam}{\lambda}
\newcommand{\al}{\alpha}

\newcommand{\ah}{\widehat{\alpha}}
\newcommand{\bh}{\widehat{\beta}}
\newcommand{\matclust}{\text{MatClust}}
\newcommand{\poisson}{\text{Poisson}}

\newcommand{\Toc}{\text{Toc}}
\newcommand{\exc}{\text{exc}}
\newcommand{\self}{\text{self}}
\newcommand{\inc}{\text{inc}}

\newcommand{\midd}{\text{mid}}
\newcommand{\pasy}{p_{\text{asy}}}

\newcommand{\prand}{p_{\text{rand}}}

\DeclareMathOperator{\BIN}{BIN}

\DeclareMathOperator{\BVN}{BVN}

\theoremstyle{plain}
\newtheorem{theorem}{Theorem}[section]

\theoremstyle{definition}

\theoremstyle{remark}

\newtheorem{remark}[theorem]{Remark}

\begin{document}
\title{
Technical Report \# KU-EC-14-1:\\
Nearest Neighbor Methods for Testing Reflexivity and Species-Correspondence}
\author{Elvan Ceyhan\thanks{Department of Mathematics, Ko\c{c} University, Sar{\i}yer, 34450, Istanbul, Turkey}
%\and
%Carey E. Priebe\thanks{Department of Applied Mathematics and Statistics, The Johns Hopkins University,
%Baltimore, MD 21218}
}
\date{\today}
\maketitle

\begin{abstract}
\noindent
Nearest neighbor (NN) methods are employed for drawing inferences about
spatial patterns of points from two or more classes.
We consider Pielou's test of niche specificity which is
defined using a contingency table based on the NN relationships between the data points.
We demonstrate that Pielou's contingency table for niche specificity is actually more appropriate
for testing reflexivity in NN structure,
hence we call this table as NN reflexivity contingency table (NN-RCT) henceforth.
We also derive an asymptotic approximation for the distribution
of the entries of the NN-RCT
and consider variants of Fisher's exact test on it.
Moreover,
we introduce a new test of class- or species-correspondence inspired by
spatial niche/habitat specificity and
the associated contingency table called species-correspondence contingency table (SCCT).
We also determine the appropriate null hypotheses and
the underlying conditions appropriate for these tests.
We investigate the finite sample performance of the tests in terms of empirical size and power
by extensive Monte Carlo simulations
and the methods are illustrated on a real-life ecological data set.
\end{abstract}

\noindent
%{\scriptsize
{\small

%\vspace{1. in}

\indent
$^\star$
This research was supported by the research agency TUBITAK via Project \# 111T767
and
by the European Commission under the
Marie Curie International Outgoing Fellowship Programme
via Project \# 329370 titled PRinHDD.\\
\indent
$^*$ {\it e-mail:} elceyhan@ku.edu.tr }

\section{Introduction}
\label{sec:intro}
The spatial point patterns in natural populations (in $\R^2$ and $\R^3$)
have received considerable attention in statistical literature.
For example, two frequently studied spatial patterns
between multiple classes or species are segregation and association (\cite{dixon:NNCTEco2002}).
Among the less studied patterns are reflexivity and niche specificity.
{\em Niche} or \emph{habitat specificity} is the collection of
biotic and abiotic conditions favoring the development and
hence the existence and abundance of a species on a spatial scale (\cite{ranker:2008}).
That is,
niche specificity is the dependence of an organism on an environment (i.e., niche or habitat).
Niche specificity can be determined by
tolerance to various conditions such as
climate, exposure to light, soil and nutrient properties (\cite{lindenmayer:2005}).
The so called {\em generalist} species have more flexible niche specificity;
that is,
they spread out in irregular numbers in the available niches
and are not confined to narrow niches but are more open to radical
changes in the environment, and are more likely to survive in new environs and alien territories.
On the other hand, the {\em specialist} species have confined niches
with high adaptation to that particular type of niche (\cite{benayas:1999}).
Niche specificity can be viewed as a factor that accounts for segregation.
\cite{pielou:1961} proposed various tests based on NN relations in a two-class setting,
namely tests of segregation, symmetry, and niche specificity, and a coefficient of segregation.
In this article,
class refers to ``species" or any other characteristic of the subjects such as sex, livelihood status and so on.
We use the NN relationships
for testing spatial patterns of NN reflexivity and species-correspondence.
Pielou's test for niche specificity for the two class case
is based on the cross-tabulation of the points
with respect to NN reflexivity and pair type.
The resultant categories are
(\emph{self,reflexive}), (\emph{mixed,reflexive}), (\emph{self,non-reflexive}),
and (\emph{mixed,non-reflexive}) pairs.
Here ``self" refers to the pair of NNs being from the same class,
and ``mixed" refers to the pair of NNs being from different classes.
Hence the NN reflexivity patterns are of two types: self or mixed NN reflexivity.
In self(resp. mixed)-NN reflexivity pattern, self(resp. mixed)-reflexive pairs are more frequent.
The reflexivity pattern is referred to as ``NN reflexivity" or ``reflexivity in NN structure" henceforth.
Non-reflexive patterns are defined similarly.

There are many methods available for testing various types of spatial patterns in literature.
These spatial tests include Pielou's test of segregation (\cite{pielou:1961}),
$K$-function (\cite{ripley:2004}),
or $J$-function (\cite{lieshout:1999}),
nearest neighbor (NN) methods (\cite{dixon:EncycEnv2002}) and so on.
An extensive survey for the tests of spatial point patterns is provided by
\cite{kulldorff:2006} who categorized and compared more than 100 such tests.
These tests are for testing spatial clustering in a one-class setting
or testing segregation of points in a multi-class setting.
The null hypothesis is some type of spatial randomness
and fully specified,
but the alternatives are often not so definite,
in the sense that for most tests the
alternatives are presented so that only
deviations from the null case are of interest
as in pure significance tests of \cite{cox:1974};
only a few tests specify for an explicit alternative clustering scheme.
However, none of the numerous tests surveyed by \cite{kulldorff:2006}
are designed for testing NN reflexivity or niche specificity.
Most of the tests for multiple classes deal with presence or lack of spatial interaction
usually in the form of spatial segregation or association between the classes.
To the author's knowledge,
the NN reflexivity test provided in this article is the only method available in literature
for assessing NN reflexivity in the NN structure of point patterns.

Habitat or niche specificity
is discussed within species diversity under the title of \emph{habitat association} in literature.
For example,
in \cite{chuyong:2011},
habitat specificity of tree species in an African forest is investigated
and is tested with torus-translation tests (\cite{harms:2001}).
More specifically,
the study area is partitioned by a fine quadrat scheme
and the number of trees from each species was counted in each
quadrat for each translation of the habitat map and the original map.
Then the relative density of each species is calculated
as the ratio of density of the species in question
and the density of all species combined.
The significance of any species' habitat association to each habitat/niche is tested
based on the rank of the original map's ratio
within the ratios obtained from the torus-translation procedure in a habitat for each species.
A similar methodology is used to determine species-habitat
associations in a subtropical forest in China (\cite{jiangshan:2009}).
Habitat association is also studied for animal species by \cite{ramsey:1994},
where adaptive cluster sampling is employed to estimate the population totals in a study region.
Tree habitat association is shown to be highly related to the genetic
structure (in the form of molecular phylogeny generated from DNA information)
in a study by \cite{pei:2011}
who assess habitat specificity by the torus-based randomization method of \cite{harms:2001}.
Habitat associations of estuarine species is investigated
with univariate and multivariate methods in \cite{hosack:2006}.
For example, in univariate analysis,
the authors analyzed the species abundance, richness,
and Shannon diversity indices of decapod and fish species
by a hierarchical longitudinal model with time as a factor and random site intercepts.
A new test called the species-correspondence test is proposed in this article.
This test is inspired by spatial niche specificity and is based on
a contingency table which is constructed using the NN relations in the data.

In this article,
we investigate the underlying assumptions for the less known
--- hence less applied compared to segregation tests ---
tests of niche specificity (due to Pielou) and NN reflexivity.
We demonstrate that the contingency table due to \cite{pielou:1961} for niche specificity
is more appropriate for reflexivity in the NN structure
(hence called \emph{NN reflexivity contingency table} (NN-RCT) in this article).
We extend Pielou's test on NN-RCT
to multi-class case together with the introduction of a new test of species-correspondence
and the associated contingency table (called species-correspondence contingency table (SCCT)).
We also suggest an approximate asymptotic distribution for the entries of the
NN-RCT for completely mapped data under random labeling (RL),
hence propose $Z$-tests for the diagonal entries
in the contingency table
and an overall $\chi^2$ NN reflexivity test
combining the $Z$-tests.
We also
investigate the use of Fisher's exact test on the NN-RCT
and determine that one of the variants has appropriate size (in rejecting the null hypothesis).
Finite sample empirical size and power comparisons are performed by Monte Carlo simulations.
We adopt the convention that
random variables and quantities are denoted with upper case letters,
while fixed quantities are denoted with lower case letters.

We describe and discuss the appropriate null cases for the tests in Section \ref{sec:null-case}.
We investigate the contingency table introduced by Pielou for niche specificity,
demonstrate that it is actually more appropriate for reflexivity in NN structure,
provide an approximate asymptotic distribution for the cell counts
and propose a new species-correspondence test in Section \ref{sec:tests-cont-tab}.
We discuss the variants of Fisher's exact test on the contingency tables in Section \ref{sec:fisher-exact-test},
consistency of the tests in Section \ref{sec:consistency-of-tests},
and provide an extensive empirical size and power analysis by Monte Carlo simulations in Section \ref{sec:empirical-size-power}.
We illustrate the methodology on an ecological data set in Section \ref{sec:example-data}
and provide some guidelines and discussion in Section \ref{sec:disc-and-conc}.

\section{Preliminaries}
\label{sec:prelim}
The concepts of NN reflexivity, segregation/association,
niche specificity and NN species-correspondence are related but different concepts,
and hence the corresponding null hypotheses are different.

For segregation/association alternatives,
the null case is that there is some sort of randomness in the spatial pattern
(as in random labeling (RL) or complete spatial randomness (CSR) independence).
The null case for the niche specificity is that  there is no relation
between the spatial distribution of a class/species and its niche or habitat,
and the null case for NN reflexivity is that values
of self- and mixed-reflexive pairs are as expected under RL or CSR independence
(these expected values will be explicitly provided in Section \ref{sec:tests-cont-tab} below).
Pielou's contingency table for niche specificity
is in fact more appropriate for testing the spatial pattern of NN reflexivity.
A base-NN pair is \emph{self-reflexive},
if both members of the pair are from the same class and each member is the NN
of the other.
Self reflexivity in NN structure is a factor that might account for or cause segregation between species.
Niche specificity and self-reflexivity in NN structure are not mutually exclusive,
in the sense that they can coexist under a segregation pattern.
On the other hand, mixed-reflexivity in NN structure might cause association in the form of,
e.g., mutualistic symbiosis between the species.
Symbiosis is an interaction between species in which there is
a close physical contact during most of lives of both participants in the form of
physiological connection or integration.
Note that the definition makes no statement about direction of interaction,
which may be mutualistic, parasitic, or commensalistic.
See, e.g., \cite{freeman:2002}.
Each type of symbiosis can be viewed as a factor causing association between the species.
The pattern of NN species-correspondence is inspired by niche specificity
and it can be viewed as a restricted form of niche specificity to
spatial proximity of the conspecifics.

The above null hypotheses can result from a more general setting.
In particular, these null cases follow provided that
there is a randomness in the NN structure in such a way that
probability of a NN of a point being from a class is proportional to the relative frequency of that class.
This assumption holds, e.g.,
under RL or CSR independence of the points from each class.
Under CSR independence,
the points from each class are independently uniformly distributed in the region of interest conditioned on the class sizes.
That is,
the points from each class are independent realizations of Homogeneous Poisson Process (HPP) with fixed class sizes
(i.e., from a binomial process).
On the other hand,
under RL,
class labels are independently and randomly assigned to a set of given locations
which could be a realization from any pattern such as HPP or some clustered or regular pattern.
Both CSR independence and RL patterns imply self- or mixed-reflexivity in NN structure and species-correspondence
exist at the expected levels (explicit forms provided below in Section \ref{sec:tests-cont-tab}).

In a two-class setting,
we label the classes as $X$ and $Y$ (or interchangeably 1 and 2, respectively).
Let $\X_{n_1}$ be a data set of size $n_1$ from class $X$
and
$\Y_{n_2}$ be a data set of size $n_2$ from class $Y$.
Then under CSR independence
we have
$\X_{n_1}=\{X_1,X_2,\ldots,X_{n_1}\}$
and
$\Y_{n_2}=\{Y_1,Y_2,\ldots,Y_{n_2}\}$
are independent and are both random samples from $\U(S)$
where
$\U$ stands for the uniform distribution on the common support
$S \subset \mathbb R^d$ for classes $X$ and $Y$.
Unless stated otherwise,
for simplicity and practical purposes,
we take $d=2$ in this article.
We combine $\X_{n_1}$ and $\Y_{n_2}$
into one data set $\mZ_n=\X_{n_1}\cup\Y_{n_2}=\{Z_1,Z_2,\ldots,Z_n\}$
where $n=n_1+n_2$.
In fact,
the data points in $\mZ_n$ are ordered pairs $(Z_i,l_i)$ for $i=1,2,\ldots,n$
where $l_i \in \{0,1\}$ or $\{X,Y\}$ are the class labels.
Under the RL pattern,
the class labels or marks are assigned randomly to points
whose locations are given.
The spatial pattern generating these point locations are referred to as the \emph{background pattern} henceforth.
Then $\mZ_n=\{z_1,z_2,\ldots,z_n\}$ is the given set of locations for $n$ points from the background pattern.
Then we have the pair of observations
$(z_i,L_i)$
where $L_i \in  \{1,2\}$ or $\{X,Y\}$ is the class label of the point $z_i$
for $i=1,2,\ldots,n$.
Then $n_1$ ($n_2$) of these $z_i$ points are assigned as class $X$ ($Y$) randomly;
i.e., the labels $L_i$ are 1 or $X$ with probability $n_1/n$
(2 or $Y$ with probability $n_2/n$)
independently for $i=1,2,\ldots,n$.
Notice that under CSR independence,
the randomness is in the (locations) of the points $Z_i$
and the class label is a fixed (deterministic) characteristic of the point
(hence denoted as $l_i$),
while under RL,
the locations of the points are fixed (hence denoted as $z_i$)
but the randomness is in the label $L_i$ associated with this point.

\section{Tests based on Contingency Tables}
\label{sec:tests-cont-tab}

\subsection{Underlying Frameworks for the Contingency Tables}
\label{sec:underlying-framework-CT}
In general,
a contingency table may result from two frameworks: row-wise and overall multinomial frameworks.
In the \emph{row-wise multinomial framework},
each row in a $k \times l$ contingency table is independent of other rows and
is from a multinomial distribution.
Hence, letting the entries of the contingency table be denoted as $N_{ij}$,
we have entries in row $i$ having $(N_{i1},N_{i2},\ldots,N_{il}) \sim \mathscr M(n_i,p_{i1},p_{i2},\ldots,p_{il})$
where $p_{ij}$ is the probability of an experimental unit
being from row category $i$ and column category $j$ simultaneously
and
$\mathscr M(n,p_1,p_2,\ldots,p_l)$ stands for the multinomial distribution
with $n$ independent trials and the probability
of trial resulting in category $i$ is $p_i$ with $\sum_{i=1}^l p_i=1$.
In the $2 \times 2$ contingency table,
the rows will have two entries,
so the row-wise multinomial distribution reduces to a binomial distribution.
More specifically, we would have
$N_{i1} \sim \mathcal \BIN(n_i,p_{i1})$
(or $N_{i2} \sim \mathcal \BIN(n_i,p_{i2})$)
for $i=1,2$
where $\BIN(n,p)$ stands
for the binomial distribution with $n$ independent trials with probability of success $p$.

On the other hand,
in the \emph{overall multinomial framework},
the cell counts are assumed to arise from independent multinomial trials.
That is,
for example, for a $k \times l$ contingency table,
\begin{multline}
\mathbf{N}=(N_{11},N_{12},\ldots,N_{1l},N_{21},N_{22},\ldots,N_{2l},\ldots,N_{k1},N_{k2}, \ldots,N_{kl})\sim \\
\mathscr M(n,p_{11},p_{12},\ldots,p_{1l},p_{21},p_{22},\ldots,p_{2l},\ldots,p_{k1},p_{k2}, \ldots,p_{kl}).
\end{multline}
Row-wise and overall multinomial frameworks are closely related.
Conditional on the row sums,
the overall multinomial framework reduces to the row-wise
multinomial framework.

\subsection{Tests based on NN Reflexivity Contingency Table}
\label{sec:ref-cont-tab}
In a two-class setting,
the patterns of self- and mixed-reflexivity in NN structure
can also be tested by a contingency table.
\cite{pielou:1961} constructed a $2 \times 2$ contingency table
partitioning the reflexive or non-reflexive pairs
into self or mixed pairs.
A pair of points $(p_1,p_2)$ are called a \emph{base-NN pair} if $p_2$ is a NN of $p_1$
where $p_1$ is called the base point and
$p_2$ is called the NN point.
A base-NN pair is called a {\em reflexive pair}, if the elements of the pair are NN to each other;
a {\em non-reflexive pair}, if the elements of the pair are not NN to each other;
a {\em self pair}, if the elements of the pair are from the same class;
a {\em mixed pair}, if the elements of the pair are from different classes.
Let $N_{s,r}$ be the observed number of self-reflexive pairs,
$N_{s,nr}$ be the observed number of self-nonreflexive pairs,
$N_{m,r}$ be the observed number of mixed-reflexive pairs,
and
$N_{m,nr}$ be the observed number of mixed-nonreflexive pairs.
Let also $N_i^{nn}$ be the number of NNs of point $Z_i$
and
$W_{ij}=\frac{1}{N_i^{nn}+N_j^{nn}}$.
Then
Then
$$N_{s,r}=\sum_{j \ne i, j=1}^n\sum_{i=1}^n W_{ij}
\I(\text{$Z_j$ is a NN of $Z_i$})\I(\text{$Z_i$ is a NN of $Z_j$}) \I(L_i=L_j),$$
$$N_{m,r}=\sum_{j \ne i, j=1}^n\sum_{i=1}^n W_{ij}
\I(\text{$Z_j$ is a NN of $Z_i$})\I(\text{$Z_i$ is a NN of $Z_j$}) \I(L_i \ne L_j),$$
\begin{multline*}
N_{s,nr}=\sum_{j \ne i, j=1}^n\sum_{i=1}^n W_{ij}
[\I(\text{$Z_j$ is a NN of $Z_i$})\I(\text{$Z_i$ is not a NN of $Z_j$})+\\
\I(\text{$Z_i$ is a NN of $Z_j$})\I(\text{$Z_j$ is not a NN of $Z_i$})] \I(L_i=L_j),
\end{multline*}
and
\begin{multline*}
N_{m,nr}=\sum_{j \ne i, j=1}^n\sum_{i=1}^n W_{ij}
[\I(\text{$Z_j$ is a NN of $Z_i$})\I(\text{$Z_i$ is not a NN of $Z_j$})+\\
\I(\text{$Z_i$ is a NN of $Z_j$})\I(\text{$Z_j$ is not a NN of $Z_i$})] \I(L_i \ne L_j).
\end{multline*}
If $\X_n$ is random sample from a continuous distribution in its support,
then $W_{ij}=1$ a.s., since each point has 1 NN a.s.
In particular under CSR independence,
we have $W_{ij}=1$ a.s.
With the partitioning of base-NN pairs according to NN reflexivity
and pair type as self or mixed,
we obtain a $2 \times 2$ contingency table,
called \emph{NN-RCT}.
See also Table \ref{tab:ref-con-tab}
where the column sum
$C_s$ is the number of self pairs,
and $C_m$ is the number of mixed pairs,
while the row sum
$N_r$ is the number of reflexive pairs,
and
$N_{nr}$ is the number of nonreflexive pairs.

\begin{table}[h]
\centering
\begin{tabular}{cc|cc|c}
\multicolumn{2}{c}{}& \multicolumn{2}{c}{pair type}& \\
\multicolumn{2}{c}{} &  self pairs   & mixed pairs  &   total  \\
\hline
& reflexive pairs  &    $N_{s,r}$    &    $N_{m,r}$    &   $N_r$  \\
\raisebox{1.5ex}[0pt] {NN reflexivity}
& non-reflexive pairs &    $N_{s,nr}$    &    $N_{m,nr}$    &   $N_{nr}$  \\
\hline
& total     &    $C_s$             &    $C_m$             &   $n$  \\
\end{tabular}
\caption{
\label{tab:ref-con-tab}
The contingency table for self- or mixed-reflexivity in NN structure, i.e.,
the NN-RCT.}
\end{table}

The \emph{niche} (or \emph{habitat}) of a species might have an impact on or
account for the existence of segregation.
If spatial niche specificity is operating,
among the reflexive pairs,
self pairs will be more frequent
than mixed pairs (\cite{pielou:1961}).
But this does not necessarily imply that the entries in the NN-RCT
would be significantly different from their expected frequencies under RL.
Pielou describes a test based on the NN-RCT
and claims that Pearson's usual $\chi^2$ test of independence
(hence, implicitly the corresponding one-sided tests) will
be appropriate for this test.
However,
a class can be restricted to a niche (i.e., can have niche specificity),
but still the self-reflexive pairs can be similar to the expected frequency.
Therefore her contingency table (in Table \ref{tab:ref-con-tab})
is useful to test the existence of self- or mixed-reflexivity in NN structure,
rather than niche specificity.
Both niche specificity and self-reflexivity in NN structure might account for segregation
or they might coexist under a segregation pattern.
On the other hand,
mixed-reflexivity in NN structure might account for the association between the classes.

\subsection{Pielou's Test Based on NN Reflexivity Contingency Table}
\label{sec:pielou-ref-table}
\cite{pielou:1961} uses the usual $\chi^2$-test of independence on the NN-RCT
in order to detect presence of niche specificity.
However, this test is used to detect independence between NN reflexivity of the pairs
and pair type as self or mixed.
Such independence would imply $\E[N_{s,r}]=C_s N_r/n$,
$\E[N_{m,r}]=C_m N_r/n$, $\E[N_{s,nr}]=C_s N_{nr}/n$,
and
$\E[N_{m,nr}]=C_m N_{nr}/n$.
Hence an excess of $N_{s,r}$ from its expected value above would imply a positive dependence
between a pair being reflexive and self pair,
and if $N_{s,r}$ is less than expected, it would imply a negative dependence.
The deviations of other entries from their expected values have similar interpretations.
Therefore, in Pielou's approach, the actual null hypothesis is
\begin{multline}
\label{eqn:Ho-Piel-reflex}
H_o: \text{independence between row and column labels} \\
\text{(i.e., independence of NN reflexivity and pair type as self or mixed)}.
\end{multline}

For the NN-RCT,
\cite{pielou:1961} suggests the use of
the Pearson's usual $\chi^2$ test with 1 df,
\begin{equation}
\label{eqn:Pielou-refl-CT}
\X^2_P=\frac{(N_{s,r}-C_s N_r/n)^2}{C_s N_r/n}+
\frac{(N_{m,r}-C_m N_r/n)^2}{C_m N_r/n}+
\frac{(N_{s,nr}-C_s N_{nr}/n)^2}{C_s N_{nr}/n}+
\frac{(N_{m,nr}-C_m N_{nr}/n)^2}{C_m N_{nr}/n}
\end{equation}
and $H_o$ is rejected when
$N_{s,r}>N_{m,r}$ and $p$-value based on the $\chi^2$ test is significant.
In the NN-RCT,
a trial is ``categorization of a base-NN pair in terms of NN reflexivity
and pair type as self or mixed".
Equivalently,
the independence between the NN reflexivity and pair types
can also be tested using the directional test statistic
\begin{equation}
\label{eqn:Zdir-one-sided}
Z_{dir}=\left( \frac{N_{s,r}}{N_r}-\frac{N_{s,nr}}{N_{nr}} \right)\sqrt{\frac{N_r\,N_{nr}\,n}{C_s\,C_m}}.
\end{equation}
Notice that the tests in Equations \eqref{eqn:Pielou-refl-CT} and \eqref{eqn:Zdir-one-sided}
are used to test the same hypothesis with the same underlying assumptions,
with only one difference that the former is for the two-sided alternative only,
while the latter can be used for both two- and one-sided alternatives.

Under positive (resp. negative) dependence, we expect $Z_{dir}>0$ (resp. $Z_{dir}<0$).
Under the usual row-wise multinomial framework,
for large $n$,
$Z_{dir}$ approximately has a $N(0,1)$ distribution.
Thus
for the negative dependence alternative,
$H_o$ is rejected when $Z_{dir} < z_{\al}$
where $z_{\al}$ is the $100 \al^{th}$ percentile of the standard normal distribution
and
for the positive dependence alternative,
$H_o$ is rejected when $Z_{dir} > z_{1-\al}$.
The two types of deviations from independence
are not distinguishable by the usual $\chi^2$ test.
Hence one can resort to the test statistic, $Z_{dir}$,
in Equation \eqref{eqn:Zdir-one-sided} for this purpose.
Furthermore,
under RL, row sums, i.e., the numbers of reflexive and non-reflexive pairs,
$N_r$ and $N_{nr}$, are fixed quantities (hence are denoted as $n_r$ and $n_{nr}$)
while they are random under CSR independence.

\begin{remark}
In the NN-RCT (in Table \ref{tab:ref-con-tab}),
row sums $N_r$ and $N_{nr}$ are fixed under RL,
hence row-wise multinomial framework is more appropriate (compared to the overall framework) under RL.
Under CSR independence, row sums would be random quantities,
so the overall multinomial framework would be more appropriate (compared to the row-wise framework).
However, a NN-RCT is unlikely to result from either framework.
In a NN-RCT,
a trial is the categorization of a base-NN pair with respect to NN reflexivity and pair type as self or mixed.
In general, in a $2 \times 2$ contingency table,
the entries $(N_{11},N_{12})$ and $(N_{21},N_{22})$ are assumed to be independent
and so are the individual trials
under the row-wise multinomial framework.
This assumption is invalid when the NN-RCT is based on completely mapped spatial data,
because independence between rows is violated.

A similar result holds under CSR independence with overall multinomial framework,
since independence between trials is violated.
Under CSR independence with sparse sampling,
the overall multinomial framework is able to model a NN-RCT approximately,
because of the inherent correlation between components or
entries of a multinomially distributed random variable.

In Pielou's test for the NN-RCT,
both of the above multinomial frameworks assume
that the trials are independent multinomial trials
which is violated by completely mapped data.
%However, when a trial is the categorization of a base-NN relation with respect to NN reflexivity and pair type as self or mixed,
%the assumption of independence between trials is violated.
Thus Pielou's test is influenced by deviations not only from
the null case but also
by deviations from dependence of trials.
The dependence can not merely be avoided
by random sub-sampling but
can be circumvent by an appropriate sparse sampling (\cite{diggle:1979}).
If the NN-RCT is constructed using a random sample of
labels of base-NN pairs in terms of NN reflexivity and pair type as self or mixed,
then the usual contingency table assumptions under the row-wise multinomial framework would hold.
Such a NN-RCT can be (approximately) obtained
only if a (small) subset of all the base-NN pairs obtained from the data in the study region were randomly selected,
i.e., if the data is obtained by an appropriate sparse sampling.
When the data were properly sparsely sampled,
we will assume that the NN-RCT satisfies the usual independence assumptions
in the row-wise multinomial framework henceforth.
In this framework, the explicit form of the null hypothesis is as in Equation \eqref{eqn:Ho-Piel-reflex}.
The assessment of various sparse sampling schemes for these tests
is a topic of ongoing research.
Our suggestion for Pielou's test on the NN-RCT is
that if the data is properly sparsely sampled,
then it is safe to employ it.
But if the data is completely mapped,
to remove the influence of spatial dependence
on Pielou's test on NN-RCT,
we suggest the usual Monte Carlo randomization
where class labels are randomly assigned to the given points a large number of times
and test statistics are computed,
and the $p$-value of the test is based on the rank (divided by the number of Monte Carlo replications)
of the test statistic of the original data in the sample of
test statistics obtained from the Monte Carlo randomization procedure.
$\square$
\end{remark}

\subsection{An Approximate Sampling Distribution for the NN Reflexivity Contingency Table}
\label{sec:approx-ref-table}
Under RL,
the number of reflexive and non-reflexive base-NN pairs are fixed,
hence the row sums, denoted as $n_r$ and $n_{nr}$, respectively, in the NN-RCT
are fixed quantities.
Let $p_{s,r}$ be the probability of a reflexive pair being a self pair
and $p_{m,nr}$ be the probability of a non-reflexive pair being a mixed pair.
Then, for the NN reflexivity tests, we have
\begin{equation}
\label{eqn:Ho-self-reflex}
H_o:\;\E[N_{s,r}]=n_r p_{s,r} ~~\text{ and } ~~\E[N_{m,nr}]=n_{nr} p_{m,nr}
\end{equation}
as our null hypothesis
where $p_{s,r}=\frac{\left( \sum_{i=1}^k n_i^2 \right)-n}{n(n-1)}$
so when $k = 2$, we have $p_{s,r}= \frac{n_1(n_1-1)}{n(n-1)}+\frac{n_2(n_2-1)}{n(n-1)}=\frac{n_1^2+n_2^2-n}{n(n-1)}$
and
$p_{m,nr}= \frac{\sum_{i \not= j} n_i n_j}{n(n-1)}$
so when $k=2$, we have $p_{m,nr}=\frac{2 n_1 n_2}{n(n-1)}$.
The alternative hypothesis for self-reflexivity in NN structure is
$H_a:\;\E[N_{s,r}] > n_r p_{s,r}$
and
the alternative for mixed-reflexivity in NN structure is
$H_a:\;\E[N_{m,nr}]>n_{nr} p_{m,nr}$.
On the other hand,
our extensive Monte Carlo simulations suggest that
$\Var[N_{s,r}] \approx 2\,n_r p_{s,r}(1-p_{s,r})$,
and
$\Var[N_{m,nr}] \approx n_{nr} p_{m,nr}(1-p_{m,nr})$.
Then,
$Z_{s,r}=\frac{N_{s,r}-\E[N_{s,r}]}{\sqrt{\Var[N_{s,r}]}}$
approximately has $N(0,1)$ distribution for large $n_r$,
and
$Z_{m,nr}=\frac{N_{m,nr}-\E[N_{m,nr}]}{\sqrt{\Var[N_{m,nr}]}}$
approximately has $N(0,1)$ distribution for large $n_{nr}$.
Assuming $N_{s,r}$ and $N_{m,nr}$ are independent,
it follows that
$\X^2_R=Z_{s,r}^2+Z_{m,nr}^2$ has asymptotically $\chi^2_2$ distribution
as both $n_r$ and $n_{nr}$ are tending to infinity.
However, this asymptotic distribution is only an approximate one,
since it is ignoring
the (nonzero) covariance, $\Cov[N_{s,r},N_{m,nr}]$.
But our Monte Carlo simulations suggest that
this covariance is approximately zero under CSR independence or RL.
Furthermore,
when $\X^2_R$ is significant,
it would only imply a significant deviation from the NN reflexivity structure under $H_o$.
To determine the direction of this deviation,
one can use the $Z$-tests,
$Z_{s,r}$ and $Z_{m,nr}$,
for the left- and right-sided alternatives.
%The two types of alternatives of self- or mixed-reflexivity in NN structure
%are not distinguishable by
%the usual $\chi^2$ test.
%Hence one should resort to the $Z$-tests above for this purpose.
%Furthermore, mixed-reflexivity in NN structure or mutualistic symbiosis might be viewed as a cause of association.

The test of NN reflexivity is closely related with
the test of segregation.
In case of segregation, self-reflexivity in NN structure can be viewed
as a cause of segregation.
Notice that mixed-reflexivity in NN structure requires NN reflexivity of class $i$ and $j$ points
with $i \not= j$.
In particular,
in the two-class case,
mixed-reflexivity in NN structure might imply association.
The extension of this approach for multi-class case is as in Section \ref{sec:pielou-ref-table}.

\begin{remark}
\label{rem:piel-ref-vs-new-ref}
Pielou's approach and the new approach on the NN-RCT
are testing different null and alternative hypotheses
(hence would have different rejection and acceptance regions).
In particular,
Pielou's approach is based on the usual $\chi^2$-test for the independence of NN reflexivity and pair type,
while the new NN reflexivity tests are based on the normal approximation of the entries
with their expected values under CSR independence or RL with completely mapped data.
Hence Pielou's test is appropriate when we have a random sample of labels of base-NN pairs
in terms of NN reflexivity and pair type as self or mixed.
$\square$
\end{remark}

\subsection{A New Test of Species-Correspondence}
\label{sec:new-niche-spec}
For a species to exhibit NN species-correspondence,
self base-NN pairs would be more abundant than expected under RL.
To detect such type of species-correspondence,
we construct a contingency table
where base-NN pairs are classified as self or mixed for each class.
Let $S_i$ be the number of self base-NN pairs for class $i$,
and $M_i$ be the number of mixed base-NN pairs with base point being from class $i$.
Then
 $$S_i=\sum_{j \ne i, j=1}^n\sum_{i=1}^n W_{ij}
\I(\text{$X_j$ is a NN of $X_i$})\I(L_i=L_j),$$
and
$$M_i=\sum_{j \ne i, j=1}^n\sum_{i=1}^n W_{ij}
\I(\text{$X_j$ is a NN of $X_i$})\I(L_i \ne L_j).$$
Recall that $W_{ij}=1$ a.s. for $\X_n$ a random sample from a continuous distribution.
The resulting contingency table is a $k \times 2$ contingency table
for $k$ classes with columns comprising of $S_i$ and $M_i$ values.
See also Table \ref{tab:nich-spec-contab-k} (left).
Notice that row sums are class sizes
(i.e., sum of row $i$ is $n_i$),
and sum of the first column (for self pairs) is $S = \sum_{i=1}^k S_i$
and sum of the second column (for mixed pairs) is $M= \sum_{i=1}^k M_i$.

\begin{table}[h]
\centering
\begin{tabular}{cc|cc|c}
\multicolumn{2}{c}{}& \multicolumn{2}{c}{pair type }& \\
\multicolumn{2}{c}{}&    self &  mixed   &   total  \\
\hline
& class 1 &    $S_1$    &    $M_1$    &   $n_1$  \\
& class 2 &    $S_2$    &    $M_2$    &   $n_2$  \\
\raisebox{1.5ex}[0pt] {base class}
& $\vdots$ & $\vdots$ &  $\vdots$ & $\vdots$\\
&class k &    $S_k$    &  $M_k$    &   $n_k$  \\
\hline
&total     &    $S$    &   $M$             &   $n$  \\
\end{tabular}
\hspace{.25cm}
\begin{tabular}{cc|ccc|c}
\multicolumn{2}{c}{}& \multicolumn{3}{c}{NN class}& \\
\multicolumn{2}{c}{}&    class 1 &  $\ldots$ & class $k$   &   total  \\
\hline
&class 1 &    $N_{11}$    &  $\ldots$ &   $N_{1k}$    &   $n_1$  \\
\raisebox{1.5ex}[0pt]{base class}& $\vdots$ & $\vdots$ & $\ddots$ & $\vdots$ & $\vdots$\\
&class $k$ &    $N_{k1}$    & $\ldots$ &    $N_{kk}$    &   $n_k$  \\
\hline
&total     &    $C_1$             &  $\ldots$ &    $C_k$             &   n  \\
\end{tabular}
\caption{
\label{tab:nich-spec-contab-k}
The NN SCCT (left) and the NNCT (right) for $k$ classes.}
\end{table}

The SCCT is closely related to the $k \times k$ nearest neighbor contingency table (NNCT) based on the same data.
%In fact, SCCT and NNCT are equivalent when $k=2$.
Here we provide a brief description of NNCTs (for more detail, see, e.g., \cite{ceyhan:cell2008}).
NNCTs are constructed using the NN frequencies of classes.
Let $N_i$ be the number of points from class $i$ for $i \in \{1,2,\ldots,k\}$ and $n=\sum_{i=1}^k N_i$.
If we record the class of each point and
its NN, the NN relationships fall into the following $k^2$ categories:
$$(1,1),\,(1,2),\ldots,(1,k);\,(2,1),\,(2,2),\ldots,(2,k);\ldots,(k,k)$$
where in category or cell $(i,j)$, class $i$ is called the \emph{base class},
and class $j$ is called the \emph{NN class}.
Denoting $N_{ij}$ as the observed
frequency of category $(i,j)$ for $i,j \in \{1,2,\ldots,k\}$, we
obtain the NNCT in Table \ref{tab:nich-spec-contab-k} (right).
That is,
$$N_{ij}=\sum_{j' \ne i', j'=1}^n\sum_{i'=1}^n W_{i'j'}
\I(\text{$Z_{j'}$ is a NN of $Z_{i'}$})\I(L_{i'}=i)\I(L_{j'}=j).$$
The number of self pairs for class $i$ is same as the number of base-NN pairs
with both base and NN class are from class $i$.
Hence $S_i=N_{ii}$ and $M_i=n_i-N_{ii}$.
Then under RL, we can determine the correct expected values, variances,
and asymptotic distributions of the cell counts in the SCCT.
In particular,
\begin{equation}
\label{eqn:Exp[Nii]}
\E[S_i]=\E[N_{ii}]=n_i(n_i-1)/(n-1)
\text{ and }
\E[M_i]=\E[n_i-N_{ii}]=n_i(n-n_i)/(n-1).
\end{equation}
Hence our null hypothesis for species-correspondence is
\begin{equation}
\label{eqn:Ho-niche-spec}
H_o:\E[S_i]=\E[N_{ii}]=n_i(n_i-1)/(n-1).
\end{equation}
Furthermore,
\begin{equation}
\label{eqn:var-Si}
\Var[S_i]=\Var[N_{ii}]=(n+R)p_{ii}+(2n-2R+Q)p_{iii}+(n^2-3n-Q+R)p_{iiii}-n^2p_{ii}^2
\end{equation}
and
$$\Var[M_i]=\Var[n_i-N_{ii}]=\Var[N_{ii}]=\Var[S_i].$$
In Equation \eqref{eqn:var-Si},
$p_{xx}$, $p_{xxx}$, and $p_{xxxx}$ are the
probabilities that a randomly picked pair, triplet, or quartet of
points, respectively, are the indicated class $i$ and are given by
\begin{equation}
\label{eqn:probs-pi}
p_{ii}=\frac{n_i\,(n_i-1)}{n\,(n-1)}, ~
p_{iii}=\frac{n_i\,(n_i-1)\,(n_i-2)}{n\,(n-1)\,(n-2)}, ~
p_{iiii}=\frac{n_i\,(n_i-1)\,(n_i-2)\,(n_i-3)}{n\,(n-1)\,(n-2)\,(n-3)},
\end{equation}
and $R$ is twice the number of reflexive pairs
and $Q$ is the number of points with shared  NNs,
which occurs when two or more
points share a NN.
Then $Q=2\,(Q_2+3\,Q_3+6\,Q_4+10\,Q_5+15\,Q_6)$
where $Q_l$ is the number of points that serve
as a NN to other points $l$ times.
The covariances of the cell counts in the same column can also be obtained as
$$\Cov[S_i,S_j]=\Cov(N_{ii},N_{jj})=(n^2-3n-Q+R)p_{iijj}-n^2p_{ii}p_{jj}$$
and
$$\Cov[M_i,M_j]=\Cov(n_i-N_{ii},n_j-N_{jj})=\Cov(N_{ii},N_{jj})$$
where $p_{iijj}=\frac{n_i\,(n_i-1)\,n_j\,(n_j-1)}{n\,(n-1)\,(n-2)\,(n-3)}.$
The covariance of cell counts in different columns is
\begin{equation}
%\label{eqn:Exp[Nij]}
\Cov[S_i,M_j]=
\begin{cases}
\Cov[N_{ii},n_i-N_{ii}]=-\Var[N_{ii}] & \text{if $i=j$,}\\
\Cov[N_{ii},n_j-N_{jj}]=-\Cov[N_{ii},N_{jj}] & \text{if $i \not= j$.}
\end{cases}
\end{equation}

Based on this contingency table,
one can obtain class-specific species-correspondence tests as
\begin{equation}
\label{eqn:cell-i-Z}
Z_{ii}=\frac{N_{ii}-\E[N_{ii}]}{\sqrt{\Var[N_{ii}]}}
\end{equation}
for $i=1,2,\ldots,k$.
Notice that the mixed column entries carry the same information
as the self column entries,
and they will yield the test statistic with the negative sign.
That is,
$(M_i-\E[M_i])/\sqrt{\Var[M_i]}=-Z_{ii}$ for each $i$,
hence the test statistics with mixed column entries are omitted.
For large $n_i$,
$Z_{ii}$ approximately has $N(0,1)$ distribution.

To obtain an overall species-correspondence test,
we can follow two approaches:

(i) Treat the self column as a vector
$\mathbf{N}_{I}=(N_{11},N_{22},\ldots,N_{kk})$.
So $\E[\mathbf{N}_{I}]$ is the vector of expected values of the entries of $\mathbf{N}_{I}$.
The variance-covariance matrix of $\mathbf{N}_{I}$, denoted $\Sigma_{\self}$, is the $k \times k$ matrix
with entry $(i,i)$ being $\Var[N_{ii}]$
and entry $(i,j)$ with $i \not= j$ being $\Cov[N_{ii},N_{jj}]$.

The overall species-correspondence test can be obtained similar to the
overall segregation test as described in \cite{ceyhan:cell2008}.
With the self column as the vector $\mathbf{N}_I$
\begin{equation}
\label{eqn:dix-NS-chisq-kxk-I}
\mathcal N_I=(\mathbf{N}_I-\E[\mathbf{N}_I])'\Sigma_{\self}^-(\mathbf{N}_I-\E[\mathbf{N}_I])
\end{equation}
where
$\Sigma_{\self}^-$ is a generalized
inverse of $\Sigma_{\self}$ (\cite{searle:2006}).
Since $\Sigma_{\self}$ is not rank deficient a.s.,
the generalized inverse in this case is equivalent to the usual matrix inverse.
For large $n_i$,
$\mathcal N_I$ approximately has a $\chi^2_k$ distribution.

(ii) Concatenate self and mixed columns to obtain the vector
$\mathbf{N}_{II}=(N_{11},N_{22},\ldots,N_{kk},n_1-N_{11},n_2-N_{22},\ldots,n_k-N_{kk})$.
So $\E[\mathbf{N}_{II}]$ is the vector of expected values of the entries of $\mathbf{N}_{II}$.
The variance-covariance matrix of $\mathbf{N}_{II}$, denoted $\Sigma_{II}$, is the $(2k) \times (2k)$ matrix
with four blocks as
$$\Sigma_{II}=
\left(
    \begin{array}{cc}
      \Sigma_{\self} & -\Sigma_{\self}\\
      -\Sigma_{\self}^t & \Sigma_{\self} \\
    \end{array}
  \right).
$$
%where $Cov$ is the variance-covariance matrix for $\mathbf{N}_{I}$.

Similarly,
with columns concatenated as the vector $\mathbf{N}_{II}$
\begin{equation}
\label{eqn:dix-NS-chisq-kxk-II}
\mathcal N_{II}=(\mathbf{N}_{II}-\E[\mathbf{N}_{II}])'\Sigma_{II}^-(\mathbf{N}_{II}-\E[\mathbf{N}_{II}])
\end{equation}
the variance-covariance matrix $\Sigma_{II}$ is rank deficient,
since the rank of this $(2k) \times (2k)$ matrix is also $k$,
we need the generalized inverse of $\Sigma_{II}$.
Then for large $n_i$,
$\mathcal N_{II}$ approximately has a $\chi^2_k$ distribution.
However,
the generalized inverse of $\Sigma_{II}$ is highly unstable,
since the covariance matrix is severely rank deficient,
hence computationally,
$\mathcal N_{II}$ might exhibit unexpected behavior
(e.g., it occasionally yields a negative test statistic
due to computational problems caused by rank deficiency,
which would not have been possible for a quadratic form as in Equation \eqref{eqn:dix-NS-chisq-kxk-II}).
Thus,
we recommend the first form of the test statistic, $\mathcal N_{I}$,
and omit $\mathcal N_{II}$ in our further discussion.
Furthermore,
when $\mathcal N_I$ is significant,
it implies the presence of significant deviation
from the species-correspondence expected under $H_o$ in Equation \eqref{eqn:Ho-niche-spec}.
But this deviation could be toward significant species-correspondence for a class,
and toward significant lack of species-correspondence for another class.
To determine the direction of deviation for each class (after a significant $\mathcal N_{I}$)
one can perform the one-sided versions of the cell-specific $Z$-tests in Equation \eqref{eqn:cell-i-Z}.

Notice also that
for $k=2$ classes,
overall species-correspondence test is equivalent to the overall test of segregation of \cite{dixon:1994}
since the SCCT and NNCT convey the same information
and both overall tests are based on $N_{11}$ and $N_{22}$ only.
In particular,
$N_{11}$ and $N_{22}$ constitute the first column of the SCCT
and
$N_{12}$ and $N_{21}$ constitute the second column of the SCCT.
But for $k>2$
the NNCT and SCCT contain different information
and
the overall species-correspondence test depends on $S_i=N_{ii}$ values only,
while the overall segregation test depends on all $N_{ij}$ values.

\section{Fisher's Exact Test for the NN Reflexivity Contingency Table}
\label{sec:fisher-exact-test}
Fisher's exact test is frequently used for contingency tables
with small cell counts and marginal sums (see \cite{agresti:1992}).
We can apply Fisher's exact test
for the $2 \times 2$ NN-RCT given in Table \ref{tab:ref-con-tab}
for the tests of NN reflexivity.
The use of exact tests on NNCTs is discussed in \cite{ceyhan:exact-NNCT}.

Fisher's exact test is feasible only for contingency tables
of small size for manual calculations.
The underlying assumption for Fisher's exact test
is that the row and column sums and grand sum are fixed,
which renders Fisher's exact test to be {\em conditional on the marginals}.
For $k \times l$ contingency tables with $\min(k,l)>2$ Fisher's exact test is two-sided only,
while for $k=l=2$, one-sided or two-sided versions are available.
In a $2 \times 2$ contingency table,
let $N_{ij}$ be the cell count for cell $(i,j)$,
$N_i$ be the sum of row $i$, and $C_j$ be the sum for column $j$.
Given the marginals (i.e., row and column sums),
$N_{11}$ determines the other three cell counts
and has the \emph{hypergeometric distribution}
with non-centrality parameter $\theta$.
In general, the null assumption of independence in the contingency tables
is equivalent to having $H_o: \theta=1$,
where $\theta$ is the non-centrality parameter (or odds ratio)
in contingency tables (\cite{agresti:1992}).

There are numerous ways to obtain $p$-values for the one-sided and two-sided
alternatives for exact inference on contingency tables (\cite{agresti:1992}).
The $p$-values based on Fisher's exact tests tend to be more conservative than most
approximate (asymptotic) ones (\cite{agresti:1992}).
Since we are more interested in the one-sided tests on the $2 \times 2$ NN-RCT,
we only consider one-sided vers2ions of Fisher's exact test.
These \emph{variants of Fisher's exact test} are described below.

\subsection{Variants of Fisher's Exact Test for the One-Sided Alternatives}
\label{sec:fisher-exact-one-sided}
To find the $p$-values for Fisher's exact test,
we find the probabilities of the contingency tables obtained from the distribution with the same row and column marginal sums.
For the one-sided alternatives,
the probabilities of more extreme tables are summed up,
excluding or including the
probability of the table itself (or some middle way).
For testing against the one-sided alternative $H_a:\,\theta>1$,
the following four methods can be obtained in computing the $p$-value.
In a $2\times 2$ contingency table,
let entry in cell $(1,1)$ be $t$,
row sums be $n_1$ and $n_2$ for the first and second rows, respectively,
and sum of column 1 be $c_1$.
Then the probability of this contingency table under $H_o$ is $p=f(t|n_1,n_2,c_1;\theta=1)$.
In the current table,
entry $(1,1)$ is $n_{11}$,
so the probability of the current table is
$p_t=f(n_{11}|n_1,n_2,c_1;\theta=1).$
For summing the $p$-values of more extreme tables than the current table,
the following variants of the exact test are obtained.
The $p$-value is calculated as $p=\sum_{t \in S} f(t|n_1,n_2,c_1;\theta=1)$ for the appropriate choice of $S$ as follows.

\begin{itemize}
\item [(i)] \emph{table-inclusive version}, $p^>_{\inc}$ with $S=\{t:\,t \geq n_{11}\}$,
\item [(ii)] \emph{table-exclusive version}, $p^>_{\exc}$ with $S=\{t:\,t> n_{11}\}$,
\item [(iii)] \emph{mid-$p$ version}, $p^>_{\midd}$ with $p=p^>_{\exc}+p_t/2$,
\item [(iv)] \emph{Tocher corrected version} which is denoted as $p^>_{\Toc}$.
\end{itemize}
Tocher's correction makes Fisher's exact
test less conservative, by including the probability for the current
table based on a randomized test (\cite{tocher:1950}).
When table-exclusive version, $p_{\exc}$, is less than the level of the test $\alpha$,
but table-inclusive version of the $p$-value, $p_{\inc}$, is larger than $\alpha$,
a random number, $U$, is generated from uniform distribution in $(0,1)$,
and if $U \ge (\alpha-p_{\exc})/p_t$, $p_{\inc}$ is used,
otherwise $p_{\exc}$ is used as the $p$-value.
That is,
\begin{equation}
\label{eqn:Tocher}
p_{\Toc}=
\begin{cases}
p_{\inc}   & \text{if $U \ge (\alpha-p_{\exc})/p_t$,}
\vspace{0.1 cm}\\
p_{\exc}   & \text{otherwise.}
\end{cases}
\end{equation}

Note also that $p^>_{\exc}=p^>_{\inc}-p_t$ and $p^>_{\midd}=p^>_{\inc}-p_t/2$.
Furthermore, $p^>_{\exc} \le p^>_{\Toc} \le p^>_{\inc}$ and $p^>_{\exc} < p^>_{\midd} < p^>_{\inc}$.

For testing against the left-sided alternative $H_a:\,\theta<1$,
the $p$-values are as above, with the inequalities being reversed.
That is,
the corresponding $p$-values are denoted as
$p^<_{\inc},\,p^<_{\exc},\,p^<_{\midd},$ and $p^<_{\Toc}$, respectively.

\subsection{Extension of the Tests to the Multi-Class Case with $k>2$}
\label{sec:multi-class-k>2}
The extension of the NN-RCT to multi-class case is straightforward,
since any multi-class data (with $k>2$) set can be categorized into the four groups
as in Table \ref{tab:ref-con-tab}
based on the relation between reflexiveness and pair type (self or mixed).
That is, the NN-RCT
can also be obtained for $k \ge 2$ classes.
Hence this contingency table is of dimension $2 \times 2$
regardless of the value of the number of classes, $k$.
However,
although the dimension is same for any number of classes,
the distribution of the column sums $(C_s,C_m)$ depends on the value of $k$.
In particular,
if $k$ gets larger,
the likelihood of reflexive NN pairs being mixed increases
and hence $C_m$ tends to increase with increasing $k$.
But this will not confound the expected cell counts in the contingency table,
since the expected values of the cell counts take into account
the row and column sums
(in Pielou's approach).
Thus a test of deviation from the expected cell counts in the NN-RCT
would not be (severely) affected by the number of classes
in the multi-class case.

In the multi-class case with $k >2$,
we recommend the following strategy:
First perform an overall omnibus test (as in ANOVA $F$-test for multi-group comparisons)
and then if the omnibus test is significant,
then perform post-hoc tests to determine the specifics of the differences.
These post-hoc tests could be pairwise tests (as in the pairwise $t$-tests)
or
one-vs-rest tests,
where one class is compared with respect to all other classes combined.
More specifically,
with $k > 2$ classes,
in the pairwise comparison, we only restrict our attention to two
classes, $i,j$ with $i\not=j$, at a time, and treat the classes as in the two-class case.
In the one-vs-rest type of test for class $i$,
we pool the remaining classes and treat them as
the other class in a two-class setting,
hence the name {\em one-vs-rest test}.
In a multi-class setting with $k$ classes,
there are $k$ one-vs-rest
type tests and $k(k-1)/2$ pairwise tests.
As $k$ increases, the first version is
computationally less intensive and easier to interpret.

The NN-RCT is still of dimension $2 \times 2$ in the case of $k >2$ classes.
But in this case, we might be interested in a break-down of the comparisons
for each pair of classes or one class compared to the rest of the classes as well.
When the test of NN reflexivity is rejected,
one might be interested in which pair of classes show self-reflexivity in NN structure
compared to others
or which classes show mixed-nonreflexivity.
Here the overall test is performed on the NN-RCT
constructed with all $k$-classes,
while pairwise test is performed with the NN-RCT
for the two classes in question,
and one-vs-rest test for, e.g., class $i$ versus rest
is performed with the NN-RCT
with two classes where one is class $i$ and the other class is taken as the remaining classes.
For any number of classes or type of post-hoc test,
the NN-RCT is of dimension $2 \times 2$,
self- and mixed-reflexivity in NN structure relations are defined with different class types.

In the multi-class case,
for the species-correspondence test, $\N_I$,
we use the $k \times 2$ SCCT
for the overall test,
while for the post-hoc tests
we need to construct $2 \times 2$ SCCTs according to the type of the test.
For pairwise tests,
we assess the species-correspondence for the two classes in question,
and for one-vs-rest type test,
we assess the species-correspondence for a class with respect to the other classes.
Furthermore,
for the pairwise tests, there is unrestricted pairwise test,
for which we extract the rows for the classes in question from the SCCT
and use the entire data in our calculation of the test statistics
(for example, we use the entire data to compute $Q$ and $R$, which are required to find the sampling
distribution of the test statistics).
On the other hand,
if interest is on the marginal interaction of two classes only,
we can construct a $2\times 2$ SCCT
based on the two classes in question, ignoring the remaining classes
(which yields restricted pairwise species-correspondence analysis).
In this case,
only the points from the two classes are used for computing all relevant quantities
such as $Q$ and $R$ for tests of species-correspondence.

In all the above cases,
the post-hoc tests can give different and seemingly conflicting results
(e.g., one class can exhibit self-reflexivity in NN structure with respect to some other class,
while mixed-reflexivity in NN structure with respect to another class.
Thus extra care should be taken which post-hoc test is used and how it should be interpreted.

\section{Consistency of Tests}
\label{sec:consistency-of-tests}
The null hypotheses are different for the
NN reflexivity and species-correspondence tests
and so are the alternative hypotheses.
Hence the comparison of the tests is inappropriate even for large samples;
but a reasonable test should have more power as the sample size increases.
So, we prove the consistency the tests in question under appropriate hypotheses.
Let $\chi^2_{\nu}(\al)$ be the $100\al^{th}$ percentile of $\chi^2$ distribution
with $\nu$ degrees of freedom.

\begin{theorem}
\label{thm:cons-piel-ref}
Let the NN-RCT be constructed by a random sample of labels of base-NN pairs
in terms of NN reflexivity and pair type as self or mixed
(or data is obtained by an appropriate sparse sampling) under a row-wise multinomial framework.
Then,
Pielou's test for the NN-RCT; i.e.,
the test rejecting independence in the NN-RCT
for $\X_P^2>\chi^2_1(1-\al)$ is consistent
where $\X_P^2$ is Pearson's test of independence given in Equation \eqref{eqn:Pielou-refl-CT}.
The one-sided tests (hence the two-sided test) using $Z_{dir}$ given in Equation \eqref{eqn:Zdir-one-sided}
are also consistent.
\end{theorem}

\noindent {\bf Proof:}
Under the null hypothesis of independence,
we have $Z_{dir} \sim N(0,1)$ for large $n$
and $Z_{dir}$ also has a normal distribution under the alternative hypothesis.
Under $H_o$, $\E[Z_{dir}]=0$
and under $H_a$, $\E[Z_{dir}|H_a]=\ve > 0$ or $\E[Z_{dir}|H_a]=\ve < 0$.
Then by the standard arguments for the consistency of $Z$-tests,
the test using $Z_{dir}$ is consistent.
Furthermore, we have $Z_{dir}^2=\X_P^2$.
The $\al$-level test based on $\X_P^2$ is equivalent to
$\al$-level two-sided test based on $Z_{dir}$.
Hence the consistency of $\X_P^2$ follows as well.
$\blacksquare$

\begin{theorem}
\label{thm:cons-niche-spec}
Let the SCCT be constructed from completely mapped spatial data under RL.
Then the test of species-correspondence; i.e.,
the test rejecting $H_o$ in Equation \eqref{eqn:Ho-niche-spec}
for $\N_I>\chi^2_k(1-\al)$ with $\N_I$ as
in Equation \eqref{eqn:dix-NS-chisq-kxk-I} is consistent.
The corresponding one-sided (hence the two-sided) cell-specific tests using $Z_{ii}$ given
in Equation \eqref{eqn:cell-i-Z} are also consistent.
\end{theorem}

\noindent {\bf Proof:}
In the $k$ class case,
let $T_{n,i}=\frac{S_i/n-\E[S_i/n]}{\sqrt{\Var[S_i/n]}}=\frac{N_{ii}/n-\E[N_{ii}/n]}{\sqrt{\Var[N_{ii}/n]}}$,
then $T_{n,i}=Z_{ii}$.
Under RL,
$\E[Z_{ii}]=0$ and
$Z_{ii}=(N_{ii}-\E[N_{ii}])/\sqrt{\Var[N_{ii}]}$ are approximately
distributed as $N(0,1)$ for large $n_i$ for $i=1,2,\ldots,k$
under the null hypotheses.
Under $H_a$, we have $\E[Z_{ii}|H_a]=\ve_i > 0$ or $\E[Z_{ii}|H_a]=\ve_i < 0$
where $\ve_i$ is a parameterization of the alternative for class $i$ for $i=1,2,\ldots,k$.
Let $\vec{\ve}=(\ve_1,\ldots,\ve_k)$,
then under $H_a:\vec \ve \ne \mathbf 0$, with $\mathbf 0$ being the vector of $k$ zeros,
let $R(\vec \ve)$ and $Q(\vec \ve)$ be the numbers of reflexive pairs
and shared pairs, respectively,
$p_{ii}(\ve_i)$, $p_{iii}(\ve_i)$, and $p_{iiii}(\ve_i)$
be the counterparts of
$p_{ii}$, $p_{iii}$, and $p_{iiii}$ in Equation \eqref{eqn:probs-pi}.
Then under $H_a$
$\Var[N_{ii}/n]=(1/n+R(\vec \ve)/n^2)p_{ii}(\ve_i)+(2/n-2R(\vec \ve)/n^2+Q(\vec \ve)/n^2)p_{iii}(\ve_i)+
(1-3/n-Q(\vec \ve)/n^2+R(\vec \ve)/n^2)p_{iiii}(\ve_i)-(p_{ii}(\ve_i))^2$.
So, under $H_a$,
$\Var[N_{ii}/n] \rightarrow 0$
as $n_i \rightarrow \infty$.
Hence the test using $Z_{ii}$ is consistent.
Also let $\lambda(\vec \ve)$ be the non-centrality parameter of $\chi^2_k$ distribution
for $\N_S$ under $H_a$.
The $\al$-level test based on $\N_S$ is consistent
since $\N_S$ is a quadratic form based on $Z_{ii}$ values,
i.e.,
$\N_S \sim \chi^2_k(\lambda(\vec \ve))$ for some $\lambda(\vec \ve)>0$.
Furthermore,
for large $n$,
the null and alternative hypotheses are equivalent to
$H_o:\,\lam=0$ versus $H_a:\,\lam=\lam(\vec \ve)>0$.
Then by standard arguments for the consistency of $\chi^2$ tests,
consistency follows.
$\blacksquare$

\begin{remark}
\label{rem:piel-consist}
The consistency result for Pielou's test on the NN-RCT is
only for sparsely sampled data under the row-wise multinomial framework.
For completely mapped spatial data,
these tests do not have the appropriate size.
In particular,
Monte Carlo simulations suggest that
the tests in Equations \eqref{eqn:Pielou-refl-CT} and \eqref{eqn:Zdir-one-sided} are liberal.
See also Section \ref{sec:empirical-size-power}.
Moreover,
the test statistics $Z_{s,r}$, $Z_{m,nr}$, and $\X^2_R$ are only approximations
and their correct (asymptotic) sampling distributions are not available,
hence consistency of these tests are omitted.
$\square$
\end{remark}

\section{Empirical Size and Power Analysis}
\label{sec:empirical-size-power}
In this section
we investigate the finite sample behavior
of the tests under their appropriate null hypotheses
and under various alternatives via Monte Carlo simulations.

\subsection{Empirical Size Analysis}
\label{sec:empirical-size}
To determine empirical size performance of the tests,
we use CSR independence and RL as our null hypotheses.
Under these patterns,
self- or mixed-reflexivity in NN structure and species-correspondence
would not deviate significantly from their expected behavior.
That is,
under these null cases,
the species-correspondence or NN reflexivity would occur at expected levels.
More specifically,
we expect that
$\E[N_{s,r}]=n_r p_{s,r}$ and $\E[N_{m,nr}]=n_{nr} p_{m,nr}$
in Equation \eqref{eqn:Ho-self-reflex} would hold for reflexivity in NN structure
and $\E[S_i]=n_i(n_i-1)/(n-1)$ in Equation \eqref{eqn:Exp[Nii]}
would hold for NN species-correspondence.

We estimate the empirical levels based on the asymptotic critical values (except for the exact tests).
For example,
let $T$ be a test with a $\chi^2_{df}$ distribution asymptotically,
and let $T_i$ be the value of test statistic for the sample generated at $i^{th}$
Monte Carlo replication for $i=1,2,\ldots,N_{mc}$.
Then the empirical size of $T$ at level $\al=0.05$, denoted $\widehat{\al}_T$ is computed as
$\widehat{\al}_T=\frac{1}{N_{mc}}\sum_{i=1}^{N_{mc}}\I(T_i \ge \chi^2_{df}(0.95))$.
Furthermore, let $Z$ be a test with a $N(0,1)$ asymptotic distribution,
and
let $Z_i$ be the value of test statistic for $i^{th}$ sample generated.
Then the empirical size of $Z$ for the left-sided alternative at level $\al=0.05$,
denoted $\widehat{\al}_Z$ is computed as
$\widehat{\al}_Z=\frac{1}{N_{mc}}\sum_{i=1}^{N_{mc}}\I(Z_i \le z_{0.05}=-1.645)$.
The empirical size for the right-sided alternative is computed as
$\widehat{\al}_Z=\frac{1}{N_{mc}}\sum_{i=1}^{N_{mc}}\I(Z_i \ge z_{0.95}=1.645)$.
For an exact test,
let $p_i$ be the $p$-value for $i^{th}$ sample generated.
Then the empirical size of this test, denoted $\widehat{\al}_e$, is computed as
$\widehat{\al}_e=\frac{1}{N_{mc}}\sum_{i=1}^{N_{mc}}\I(p_i \le 0.05)$.

\subsubsection{Empirical Size Analysis under CSR Independence}
\label{sec:empirical-size-CSR}
We consider the two-class case,
with classes $X$ and $Y$ (also referred as classes 1 and 2)
of sizes $n_1$ and $n_2$, respectively.
Let $\{X_1,\ldots,X_{n_1}\}$ be the set of class 1 points
and $\{Y_1,\ldots,Y_{n_2}\}$ be the set of class 2 points.
Under $H_o$,
at each of $N_{mc}=10000$ replicates,
we generate $X$ and $Y$ points independently of each other and iid from $\U((0,1)\times (0,1))$,
the uniform distribution on the unit square.
We consider two cases for CSR independence:
\begin{itemize}
\item[] \textbf{Case 1}:
We generate $n_1=n_2=n =10,20,30,40,50$ points
iid from $\U((0,1)\times (0,1))$.
In this case,
the sample sizes are equal and increasing.
\item[] \textbf{Case 2}:
To determine the influence of differences in the sample sizes
(i.e., differences in relative abundances of classes) on the empirical levels of the tests,
we generate the samples from the CSR independence pattern with $n_1=20$ and $n_2=20,30,\ldots,60$.
\end{itemize}

\begin{table}[]
\centering
\begin{tabular}{|c||c|c|c||c|c|c||c|c|c|}
\hline
\multicolumn{1}{|c}{ }&\multicolumn{9}{c|}{Empirical significance levels of the tests under CSR independence} \\
\hline
\multicolumn{1}{|c}{ }&\multicolumn{9}{c|}{case 1: $n_1=n_2=n=10,20,\ldots,50$} \\
\hline
$n$  & $\ah_P$& $\ah^{>}_{dir}$ & $\ah^{<}_{dir}$ & $\ah_R$
& $\ah^Z_{s,r}$ & $\ah^Z_{m,nr}$ & $\ah_{sc}$  & $\ah^Z_{11}$ & $\ah^Z_{22}$\\
\hline
10 & .0439 & .1031 & .0580 & .0503 & .0568 & .0580 & .0432 & .0454 & .0465\\
\hline
20 & .0555 & .0823 & .0507 & .0452 & .0508 & .0507 & .0457 & .0517 & .0522\\
\hline
30 & .0564 & .0822 & .0458 & .0475 & .0484 & .0458 & .0485 & .0573 & .0493\\
\hline
40 & .0649 & .0825 & .0474 & .0484 & .0480 & .0474 & .0501 & .0507 & .0525\\
\hline
50 & .0678 & .0872 & .0525 & .0458 & .0459 & .0525 & .0472 & .0454 & .0472\\
\hline
\hline
\multicolumn{1}{|c}{ }&\multicolumn{9}{c|}{case 2: $n_1=20, n_2=20,30,\ldots,60$} \\
\hline
$n_2$ & $\ah_P$& $\ah^{>}_{dir}$ & $\ah^{<}_{dir}$ & $\ah_R$
& $\ah^Z_{s,r}$ & $\ah^Z_{m,nr}$ & $\ah_{sc}$ & $\ah^Z_{11}$ & $\ah^Z_{22}$ \\
\hline
20 & .0531 & .0833  & .0485 & .0449 & .0523 & .0485 & .0437 & .0482 & .0517\\
\hline
30 & .0566 & .0843 & .0500 & .0454 & .0447 & .0500 & .0480 & .0521 & .0479\\
\hline
40 & .0552 & .0810 & .0405 & .0337 & .0387 & .0405 & .0489 & .0313 & .0455\\
\hline
50 & .0559 & .0839 & .0399 & .0290 & .0226 & .0399 & .0427 & .0295 & .0478\\
\hline
60 & .0511 & .0764 & .0337 & .0235 & .0238 & .0337 & .0452 & .0395 & .0495\\
\hline
\end{tabular}
\caption{
\label{tab:size-CSR-cases}
The empirical significance levels of the tests under CSR independence cases 1 and 2
with $N_{mc}=10000$ at $\alpha=.05$.
$\ah_P$ is the empirical significance level for the $\chi^2$-test of independence with 1 df for the NN-RCT,
$\ah^{>}_{dir}$ (resp. $\ah^{<}_{dir}$) for the right(resp. left)-sided alternative
for the directional test, $Z_{dir}$, in Equation \eqref{eqn:Zdir-one-sided};
$\ah_R$ for the $\chi^2$ test statistic $\X^2_R$ for self- or mixed-reflexivity in NN structure;
$\ah^Z_{s,r}$ for the self-reflexivity in NN structure test statistic, $Z_{s,r}$;
$\ah^Z_{m,nr}$ for the mixed-nonreflexivity test statistic, $Z_{m,nr}$;
$\ah^Z_{11}$ and $\ah^Z_{22}$ for the cell-specific tests for cells 1 and 2 (for segregation);
and
$\ah_{sc}$ for the $\chi^2$ test of species-correspondence, $\N_I$.}
\end{table}

The empirical significance levels for the tests
under CSR independence cases 1 and 2 are
presented in Table \ref{tab:size-CSR-cases},
where
$\ah_P$ is the empirical significance level for $\X^2_P$,
Pearson's $\chi^2$-test of independence with 1 df for the NN-RCT
(suggested by Pielou),
$\ah^{>}_{dir}$ (resp. $\ah^{<}_{dir}$) is for the right(resp. left)-sided alternative,
i.e., positive (resp. negative) dependence between NN reflexivity and self pairs,
for the directional test, $Z_{dir}$, in Equation \eqref{eqn:Zdir-one-sided};
$\ah_R$ is for the $\chi^2$ test statistic, $\X^2_R$, for self- or mixed-reflexivity in NN structure;
$\ah^Z_{s,r}$ is for the self-reflexivity in NN structure test statistic, $Z_{s,r}$;
$\ah^Z_{m,nr}$ is for the mixed-nonreflexivity test statistic, $Z_{m,nr}$;
$\ah^Z_{11}$ and $\ah^Z_{22}$ are for the cell-specific tests for cells 1 and 2 (for segregation)
(see, e.g., \cite{dixon:1994} and \cite{ceyhan:cell2008} for details on the cell-specific tests);
and
$\ah_{sc}$ is for the $\chi^2$ test of species-correspondence, $\N_I$.
For $N_{mc}=10000$ replications,
an empirical size estimate is deemed conservative, if smaller than 0.0464
while it is deemed liberal, if larger than 0.0536
at .05 level (based on binomial critical values with $n=10000$ trials
and probability of success 0.05).
Under CSR independence case 1,
notice that $\ah^{>}_{dir}$ is significantly larger than 0.05
(i.e., $Z_{dir}$ is significantly liberal)
for all sample size combinations
and the $\chi^2$-test of independence for the NN-RCT
is liberal for large samples (i.e., for $n \ge 40$).
The other tests seem to be of the desired level for each sample size considered.
Under case 2,
observe that $Z_{dir}$ is liberal at .05 level
(although less liberal compared to case 1),
and contrary to case 1,
$\chi^2$-test of independence for the NN-RCT, $\X^2_P$
is about the desired level for each sample size combination.
Furthermore,
$\ah^{<}_{dir}$, $\ah^Z_{s,r}$, $\ah^Z_{m,nr}$, $\ah_R$, and $\ah^Z_{11}$
seem to be significantly less than .05 (i.e., the corresponding tests are conservative) when
the relative abundance ratio gets larger than two (i.e., when $n_2/n_1 \ge 2$).
The other tests show similar size performance as in case 1.
In both cases,
$\X^2_P$ has larger size estimates
compared to $\X^2_R$
and
the right-sided directional test $Z_{dir}$
has larger size estimates compared to $Z$-test for self-reflexivity in NN structure, $Z_{s,r}$.
On the other hand,
the left-sided directional test and  $Z$-test for mixed-nonreflexivity, $Z_{m,nr}$,
has equal size estimates.
Furthermore,
we recommend the use of the Monte Carlo randomized versions or the use of Monte Carlo critical values for
$\X^2_P$ and the right-sided alternative for $Z_{dir}$ for balanced sample sizes.
Also,
we recommend the use of the Monte Carlo randomized versions or the use of Monte Carlo critical values for
all the tests except $\X^2_P$, cell-specific test for the larger class
and the species-correspondence test, $\N_I$, for unbalanced sample sizes.
A Monte critical value is determined as the appropriately ranked value of the test statistic
in a certain number of generated data sets under the null hypothesis.
The class sizes are said to be \emph{balanced},
if the relative abundances of the classes are close to one,
and they are called \emph{unbalanced},
if the relative abundances deviate substantially from one.

\begin{table}[ht]
\centering
\begin{tabular}{|c||c|c|c|c||c|c|c|c|}
\hline
\multicolumn{9}{|c|}{Empirical significance levels for the one-sided exact tests} \\
\multicolumn{9}{|c|}{on the NN-RCT under CSR independence} \\
\hline
\multicolumn{1}{|c}{ }&\multicolumn{8}{c|}{case 1: $n_1=n_2=10,20,\ldots,50$} \\
\hline
$(n_1,n_2)$  & $\ah^S_{\inc}$ & $\ah^S_{\exc}$ & $\ah^S_{\midd}$ & $\ah^S_{\Toc}$ &
$\ah^M_{\inc}$ & $\ah^M_{\exc}$ & $\ah^M_{\midd}$ & $\ah^M_{\Toc}$ \\
\hline
$(10,10)$ & .0461 & .1817 & .0771 & .0878 & .0472 & .1710 & .0764 & .0879\\
\hline
$(20,20)$ & .0510 & .1408 & .0709 & .0795 & .0556 & .1411 & .0764 & .0826\\
\hline
$(30,30)$ & .0543 & .1255 & .0746 & .0815 & .0539 & .1297 & .0766 & .0837\\
\hline
$(40,40)$ & .0602 & .1229 & .0806 & .0839 & .0596 & .1226 & .0801 & .0855\\
\hline
$(50,50)$ & .0628 & .1163 & .0823 & .0856 & .0597 & .1105 & .0779 & .0808\\
\hline
\multicolumn{1}{|c}{ }&\multicolumn{8}{c|}{case 2: $n_1=20, n_2=20,30,\ldots,60$} \\
\hline
$(20,20)$ & .0540 & .1418 & .0761 & .0835 & .0539 & .1451 & .0762 & .0809\\
\hline
$(20,30)$ & .0572 & .1394 & .0804 & .0864 & .0532 & .1308 & .0752 & .0822\\
\hline
$(20,40)$ & .0549 & .1193 & .0735 & .0795 & .0542 & .1247 & .0742 & .0800\\
\hline
$(20,50)$ & .0535 & .1683 & .0736 & .0801 & .0432 & .1063 & .0606 & .0682\\
\hline
$(20,60)$ & .0509 & .1068 & .0687 & .0734 & .0437 & .1019 & .0610 & .0679\\
\hline
\end{tabular}
\caption{
\label{tab:exact-one-sided-CSR}
The empirical significance levels for Fisher's one-sided exact tests
on the NN-RCT under CSR independence cases 1 and 2 with
$N_{mc}=10000$, for some combinations of $n_1,n_2$ at $\alpha=.05$.
$\ah^S_{\inc}$ is the empirical significance level for the table-inclusive version of the one-sided
exact test on NN-RCT,
$\ah^S_{\exc}$ is for the table-exclusive version,
$\ah^S_{\midd}$ is for the mid-$p$-value version,
$\ah^S_{\Toc}$ is for the Tocher corrected version.
The notation is similar for the mixed-reflexivity in NN structure alternative with $S$ replaced with $M$.}
\end{table}

The empirical significance levels on the exact tests on the NN-RCT
under CSR independence cases 1 and 2 are presented in Table \ref{tab:exact-one-sided-CSR},
where $\ah^S_{\inc}$ is the empirical significance level for the table-inclusive version of the one-sided
exact test on the NN-RCT for positive dependence between NN reflexivity and self pairs,
$\ah^S_{\exc}$ is for the table-exclusive version,
$\ah^S_{\midd}$ is for the mid-$p$-value version,
$\ah^S_{\Toc}$ is for the Tocher corrected version.
The notation is similar for positive dependence between NN reflexivity and mixed pairs
(or negative dependence between NN reflexivity and self pairs)
with $S$ replaced with $M$.
Notice that only the table inclusive versions are about the desired level,
while the others are extremely liberal.
Hence in what follows, only the table inclusive versions are used for
the exact inference on NN-RCTs.

\subsubsection{Empirical Size Analysis under RL}
\label{sec:empirical-size-RL}
For the RL pattern,
we consider RL of class labels of 1 and 2 (or $X$ and $Y$) to the points which are generated from
homogeneous or clustered background patterns.
To reduce the influence of a particular background realization on the size performance of the tests,
we generate 100 different realizations of each background pattern.
At each background realization,
$n_1$ of the points are labeled as class 1 and the remaining $n_2=n-n_1$
points are labeled as class 2.

\begin{itemize}
\item[] \textbf{Types of the Background Patterns:}
\item[] \textbf{Case 1:}
The background points are generated as $Z_i \stackrel{iid}{\sim} \U((0,1)\times(0,1))$ for $i=1,2,\ldots,n$.
That is the background points, $\mZ_n$, are
generated iid uniform in the unit square $(0,1)\times(0,1)$.
We consider $n_1=n_2=10,20, \ldots,50$ to determine the effect of equal but increasing equal sample sizes.

\item[] \textbf{Case 2:}
The background points, $\mZ_n$, are generated as in case 1 above with
$n_1=20$ and $n_2=20,30,\ldots,60$ to determine the differences in the relative abundances of the classes
with number of class 1 points fixed and number of class 2 points increasing.

\item[] \textbf{Case 3:}
The background points, $\mZ_n$, are generated from a Mat\'{e}rn cluster process, $\matclust(\kappa,r,\mu)$
(\cite{baddeley:2005}).
In this process,
first ``parent" points are generated from a Poisson process with intensity $\kappa$.
Then each parent point is replaced by $N$ new points
which are generated iid inside the circle of radius $r$ centered at the parent point
with $N \sim \poisson(\mu)$.
Each background realization is a
one realization of $\mZ_n$ and is generated from $\matclust(\kappa,r,\mu)$.
Let $n$ be the number of points in a particular realization.
Then $n_1=\lfloor n/2  \rfloor$ of these points are labeled as class 1
where $\lfloor x \rfloor$ stands for the floor of $x$,
and $n_2=n-n_1$ as class 2.
In our simulations,
we use $\kappa=2,4,\ldots,10$, $\mu=\lfloor 100/\kappa  \rfloor$, and  $r=0.1$.
That is,
we take $(\kappa,\mu) \in \{(2,50), (4,25) \ldots, (10,10)\}$,
so as to have about 100 background points on the average
with about half of them being from class 1 and the other half being from class 2.
\end{itemize}

For each case, the RL scheme described is repeated $1000$ times for each
$(n_1,n_2)$ combination at each of 100 background realizations.
In RL cases 1 and 2, the points are from HPP in the unit square with fixed $n_1$ and $n_2$
(i.e., from a binomial process),
where case 1 is for assessing the effect of equal but increasing sample sizes on the tests,
while case 2 is for assessing the effect of increasing differences in sample sizes of the classes
(with one class size being fixed, while the other is increasing).
On the other hand,
in the background realizations of case 3,
centers and numbers of clusters are random.
On the average,
with increasing $\kappa$,
the cluster sizes tend to decrease and
the number of clusters tend to increase (so as to have fixed class sizes on the average).
Hence in case 3,
we investigate
the influence of increasing number of clusters with randomly determined centers
on the size performance of the tests.

The empirical size estimates of the tests
under RL cases 1-3 are presented in Table \ref{tab:size-RL-cases}.
For $N_{mc}=100000$ replications,
an empirical size estimate is deemed conservative, if smaller than 0.04887
while it is deemed liberal, if larger than 0.05113
at .05 level (based on binomial critical values with $n=10000$ trials
and probability of success 0.05).
The size performance under cases 1 and 2 are similar to that under CSR independence
cases 1 and 2, respectively.
However,
under RL case 3,
$\X^2_P$ is liberal for each $\kappa$ value,
which would be expected,
since for each $\kappa$ value $n_1 \approx n_2 \approx 50$
(and this test was liberal for this sample size under RL case 1).
Notice also that the size estimates of the tests
are not influenced by the number of clusters, $\kappa$,
when the class sizes are fixed.
The empirical size estimates of the exact tests
for the table inclusive versions of the right-sided and left-sided exact tests
on the NN-RCT
are denoted as $\ah_F^{>}$, and $\ah_F^{<}$, respectively,
for notational convenience.
The one-sided versions of Fisher's exact test on the NN-RCT
seem to be slightly liberal for larger sample sizes under RL case 1.
Also, the size estimates are not influenced by the number of clusters
provided class sizes are fixed.

\begin{table}[]
\centering
\begin{tabular}{|c||c|c|c||c|c|c||c|c|c||c|c|}
\hline
\multicolumn{1}{|c}{ }&\multicolumn{11}{c|}{Empirical significance levels of the tests under RL} \\
\hline
\multicolumn{1}{|c}{ }&\multicolumn{11}{c|}{case 1} \\
\hline
$n$  & $\ah_P$& $\ah^{>}_{dir}$ & $\ah^{<}_{dir}$ & $\ah_R$
& $\ah^Z_{s,r}$ & $\ah^Z_{m,nr}$ & $\ah_{sc}$ & $\ah^Z_{11}$ & $\ah^Z_{22}$ & $\ah_F^{>}$ & $\ah_F^{<}$ \\
\hline
10 & .04492 & .10060 & .05157 & .04902 & .05601 & .05157 & .04281 & .04513 & .04625 & .04523 & .04856\\
\hline
20 & .05617 & .08304 & .05059 & .04669 & .05407 & .05059 & .04511 & .05349 & .05209 & .05610 & .05559\\
\hline
30 & .06038 & .08353 & .04468 & .05006 & .04657 & .04468 & .04862 & .05220 & .05258 & .05561 & .05717\\
\hline
40 & .06690 & .08382 & .04768 & .04952 & .04996 & .04768 & .04782 & .05232 & .05217 & .05994 & .05963\\
\hline
50 & .06986 & .08464 & .05053 & .04864 & .05009 & .05053 & .04942 & .04740 & .04642 & .05961 & .06142\\
\hline
\multicolumn{1}{|c}{ }&\multicolumn{11}{c|}{case 2} \\
\hline
$n_2$ & $\ah_P$ & $\ah^{>}_{dir}$ & $\ah^{<}_{dir}$ & $\ah_R$
& $\ah^Z_{s,r}$ & $\ah^Z_{m,nr}$ & $\ah_{sc}$ & $\ah^Z_{11}$ & $\ah^Z_{22}$ & $\ah_F^{>}$ & $\ah_F^{<}$ \\
\hline
20 & .05581 & .08211 & .05207 & .04744 & .05437 & .05207 & .04602 & .05479 & .05414 & .05511 & .05497\\
\hline
30 & .05793 & .08594 & .04598 & .04314 & .04436 & .04598 & .04735 & .05050 & .04886 & .05331 & .05334\\
\hline
40 & .05345 & .07986 & .04169 & .03300 & .03439 & .04169 & .04551 & .03375 & .04358 & .05390 & .05014\\
\hline
50 & .05516 & .08028 & .03826 & .02896 & .02488 & .03826 & .04611 & .03456 & .04893 & .05140 & .04786\\
\hline
60 & .05048 & .07368 & .03288 & .02381 & .02238 & .03288 & .04395 & .04042 & .04749 & .04981 & .04695\\
\hline
\multicolumn{1}{|c}{ }&\multicolumn{11}{c|}{case 3} \\
\hline
$\kappa$  & $\ah_P$ & $\ah^{>}_{dir}$ & $\ah^{<}_{dir}$ & $\ah_R$
& $\ah^Z_{s,r}$ & $\ah^Z_{m,nr}$ & $\ah_{sc}$ & $\ah^Z_{11}$ & $\ah^Z_{22}$ & $\ah_F^{>}$ & $\ah_F^{<}$ \\
\hline
2 & .06698 & .08511 & .04860 & .04886 & .05125 & .04860 & .04713 & .04883 & .04835 & .05817 & .05795\\
\hline
4 & .06731 & .08451 & .05087 & .04866 & .04895 & .05087 & .04665 & .04858 & .04911 & .05763 & .05895\\
\hline
6 & .06656 & .08529 & .05108 & .04963 & .05191 & .05108 & .04935 & .05003 & .05070 & .05988 & .05771\\
\hline
8 & .06787 & .08346 & .04829 & .04777 & .04858 & .04829 & .04749 & .04839 & .04862 & .05966 & .06023\\
\hline
10 & .06709 & .08311 & .05056 & .04949 & .04913 & .05056 & .04858 & .05022 & .04972 & .05981 & .05887\\
\hline
\end{tabular}
\caption{
\label{tab:size-RL-cases}
The empirical significance levels of the tests under RL cases 1-3 with $N_{mc}=1000$
for each of 100 background realization at $\alpha=.05$.
$\ah_F^{>}$ and $\ah_F^{<}$ stand for
for Fisher's exact test (table inclusive versions) on the NN-RCT
for the right-sided and left-sided alternatives.
The empirical size labeling for other tests is as in Table \ref{tab:size-CSR-cases}.}
\end{table}

Based on the empirical size performance of the tests under CSR independence and RL,
we conclude that directional test, $Z_{dir}$, is liberal for the right-sided alternative for small to large samples
and Pielou's $\chi^2$ test of independence on the NN-RCT, $\X^2_P$, is liberal for large samples.
So we recommend Monte Carlo randomization for these tests under these situations.
Furthermore,
left-sided directional $Z$-test and new $Z$-tests for self- and mixed-reflexivity in NN structure
and $\chi^2$ test of NN reflexivity
and the cell-specific tests for the smaller class
are all conservative when the relative abundances
of the classes are very different.
That is,
these tests are severely confounded by the differences in relative abundances of the classes.
Therefore,
we recommend the use of these tests when the sample sizes are balanced
and
we recommend Monte Carlo randomization for these tests for unbalanced sample sizes.
We also observe that
the new test of species-correspondence, $\N_I$, is appropriate for balanced
or unbalanced sample sizes.
For the exact test on the NN-RCT,
we recommend the table-inclusive versions for both one-sided directions
as they have the best empirical size performance (i.e., they are closest to the nominal level).
By the virtue of exact tests,
we recommend their use (in particular those for the NN-RCT)
for small sample sizes.

\subsection{Empirical Power Analysis}
\label{sec:empirical-power}
To compare the empirical power performance of the tests,
we consider various alternative cases
for self- or mixed-reflexivity in NN structure
and species-correspondence.
The empirical power estimates are computed similar to
the size estimates in Section \ref{sec:empirical-size}.

\textbf{Case I:}
For the first class of alternatives,
we generate
$X_i\stackrel{iid}{\sim} \U((0,1)\times(0,1))$ for $i=1,\ldots,n_1$
and
$Y_j \stackrel{iid}{\sim} \BVN(1/2,1/2,\sigma_1,\sigma_2,\rho)$ for $j=1,\ldots,n_2$,
where $\BVN(\mu_1,\mu_2,\sigma_1,\sigma_2,\rho)$
is the bivariate normal distribution with mean $(\mu_1,\mu_2)$
and covariance
$ \left[ \begin {array}{cc}
\sigma_1 &\rho\\
\noalign{\medskip}
\rho &\sigma_2
\end {array} \right].
$
In our simulations,
we set $\sigma_1=\sigma_2=\sigma$ and $\rho=0$.
We consider the following three alternatives:
\begin{equation}
\label{eqn:Ha-case-I}
H_{I}^{1}: \sigma=1/10,\;\;\;H_{I}^{2}: \sigma=1/20, \text{ and } H_{I}^{3}: \sigma=1/30.
\end{equation}
The classes 1 and 2 (i.e., $X$ and $Y$) have different distributions with different local intensities.
In particular,
$X$ points are a realization of uniform distribution in the unit square,
while $Y$ points are clustered around the center of the unit square $(1/2,1/2)$
with the level of clustering increasing as $\sigma$ decreases.
This suggests a high level of species-correspondence for $Y$ points around the center of the unit square
compared to $X$ points,
which in turn implies segregation of $Y$ points from $X$ points.

\begin{table}[ht]
\centering
\begin{tabular}{|c|c|c||c|c|c||c|c|c|}
\hline
\multicolumn{1}{|c}{ }&\multicolumn{8}{|c|}{Power estimates under the case I alternatives} \\
\hline
& $\bh_F^{>}$ & $\bh_F^{<}$ & $\bh_R$ &
$\bh^{Z,>}_{s,r}$ & $\bh^{Z,<}_{m,nr}$ & $\bh_{sc}$ & $\bh^{Z,>}_{11}$ & $\bh^{Z,>}_{22}$ \\
\hline
$H_{I}^1$ & .0805 & .0203 & .9935 & .9743 & .9651 & .9929 & .9887 & .9972\\
\hline
$H_{I}^2$ & .1262 & .0032 & 1.000 & 1.000 & .9996 & 1.000 & 1.000 & 1.000\\
\hline
$H_{I}^3$ & .1337 & .0005 & 1.000 & 1.000 & 1.000 & 1.000 & 1.000 & 1.000 \\
\hline
\end{tabular}
\caption{
\label{tab:alternative-1}
The power estimates under the case I alternatives in Equation \eqref{eqn:Ha-case-I} with $N_{mc}=10000$,
$n_1=n_2=40$ at $\alpha=.05$.
$\bh_F^{>}$ and $\bh_F^{<}$ are power estimates for the exact tests on the NN-RCT;
$\bh_R$ for the $\chi^2$ test statistic, $\X^2_R$, for self- or mixed-reflexivity in NN structure;
$\bh^Z_{sr}$ for the self-reflexivity in NN structure test statistic, $Z_{s,r}$;
$\bh^Z_{mn}$ for the mixed-nonreflexivity test statistic, $Z_{m,nr}$;
$\bh_{sc}$ for the $\chi^2$ test of species-correspondence $\N_I$;
$\bh^Z_{11}$ and $\bh^Z_{22}$ for the cell-specific tests for cells 1 and 2 (for segregation).
The ``$>$" (``$<$") sign in the superscript implies the power is estimated for the right-sided (left-sided) alternative.}
\end{table}

The empirical power estimates under the alternatives,
$H_{I}^{1}-H_{I}^{3}$ with $n_1=n_2=40$
are presented in Table \ref{tab:alternative-1},
where
$\bh_F^{>}$ and $\bh_F^{<}$ are for the right-sided and
left-sided exact tests on the NN-RCT, respectively;
$\bh_R$ is for the $\chi^2$ test statistic, $\X^2_R$, for self- or mixed-reflexivity in NN structure;
$\bh^Z_{sr}$ is for the self-reflexivity in NN structure test statistic, $Z_{s,r}$;
$\bh^Z_{mn}$ is for the mixed-nonreflexivity test statistic, $Z_{m,nr}$;
$\bh_{sc}$ is for the $\chi^2$ test of species-correspondence, $\N_I$;
$\bh^Z_{11}$ and $\bh^Z_{22}$ are for the cell-specific tests for cells 1 and 2 (for segregation).
\emph{We omit the power estimates for the $\chi^2$-test of independence and
one-sided directional tests on the NN-RCT,
since they are undefined when an entire column of the NN-RCT is zero,
which happens with non-negligible probability under case I alternatives.}
Under the case I alternatives,
the exact tests for the right-sided alternative on the NN-RCT indicates a slight power
for self-reflexivity in NN structure, while the exact test for the left-sided alternative has virtually zero power.
On the other hand,
$\chi^2$ NN reflexivity test, and right-sided test of self-reflexivity in NN structure, $Z_{s,r}$
and left-sided mixed-nonreflexivity test, $Z_{m,nr}$ have very high power estimates about 1.000,
which implies a high degree of self-reflexivity in NN structure,
which is caused by the high level of clustering of $Y$ points around the center of the unit square.
Similarly,
the species-correspondence test, $\N_I$, and the right-sided cell-specific tests for cells $(1,1)$ and $(2,2)$
are highly significant,
which indicates the high level of segregation of $Y$ points from $X$ points.
Notice that the $\chi^2$ and $Z$-tests for self-reflexivity in NN structure have higher power compared to the exact tests,
and the species-correspondence and cell-specific segregation tests have the highest power estimates.

\textbf{Case II:}
For the second type of alternative,
first, we generate $X_i\stackrel{iid}{\sim} \U((0,1)\times(0,1))$ for
$i=1,2,\ldots,n_1$
and for each $j=1,2,\ldots,n_2$, we generate $Y_j$ around a
randomly picked $X_i$ with probability $p$ in such a way that
$Y_j=X_i+R_j\,(\cos T_j, \sin T_j)^t$
where $v^t$ stands for transpose of the vector $v$,
$R_j \sim \U(0,\min_{i \not=j}d(X_i,X_j))$ and $T_j \sim \U(0,2\,\pi)$
or generate $Y_j$ uniformly in the unit square with probability $1-p$.
In the pattern generated,
$Y_j$ are more associated with $X_i$.
%That is, a $Y$ point is more likely to be the NN of an $X$ point,
%but the converse is not as likely.
%Hence the NN structure is asymmetric.
The three values of $p$ constitute the following alternatives:
\begin{equation}
\label{eqn:Ha-case-II}
H_{II}^{1}: p=.25,\;\;\;H_{II}^{2}: p=.50, \text{ and } H_{II}^{3}: p=.75.
\end{equation}

\begin{table}[ht]
\centering
\begin{tabular}{|c|c|c||c|c|c||c|c|c|}
\hline
\multicolumn{1}{|c}{ }&\multicolumn{8}{|c|}{Power estimates under the case II alternatives} \\
\hline
& $\bh_F^{>}$ & $\bh_F^{<}$ & $\bh_R$ &
$\bh^{Z,<}_{s,r}$ & $\bh^{Z,>}_{m,nr}$  & $\bh_{sc}$ & $\bh^{Z,<}_{11}$ & $\bh^{Z,<}_{22}$ \\
\hline
$H_{II}^1$  & .0003 & .3402 & .3128 & .5533 & .0677 & .1651 & .3142 & .2324\\
\hline
$H_{II}^2$  & .0001 & .6651 & .8642 & .9629 & .1121 & .5064 & .7386 & .3707\\
\hline
$H_{II}^3$  & .0000 & .8721 & .9985 & .9999 & .1268 & .7777 & .9491 & .2431\\
\hline
\end{tabular}
\caption{
\label{tab:alternative-2}
The power estimates under the case II alternatives in Equation \eqref{eqn:Ha-case-II} with $N_{mc}=10000$,
$n_1=n_2=40$ at $\alpha=.05$.
The empirical power labeling and superscripting for ``$<$" and ``$>$" are as in Table \ref{tab:alternative-1}.}
\end{table}

In this case,
$X$ points constitute a realization of the uniform distribution in the unit square,
while $Y$ points are clustered around the $X$ points,
and the level of clustering increases as $p$ increases.
The empirical power estimates under the alternatives,
$H_{II}^1-H_{II}^3$ with $n_1=n_2=40$
are presented in Table \ref{tab:alternative-2}.
Notice that
the right-sided exact test on NN-RCT has virtually zero power,
while the left-sided exact test high power which increases as $p$ increases.
$\chi^2$ NN reflexivity and species-correspondence tests have high power which increases as $p$ increases,
but $Z_{s,r}$ has high power for the left-sided alternative
and $Z_{m,nr}$ has mild power for the right-sided alternative,
which indicates significant lack of self-reflexivity in NN structure but presence of mild mixed-nonreflexivity.
The cell-specific tests have high power for the left-sided alternative
with $Z_{11}$ having higher power estimates.
Hence there is significant mixed-reflexivity in NN structure or mixed-nonreflexivity,
and significant lack of species-correspondence or significant association between the classes.
$Z_{11}$ having higher power for the left-sided alternative is due to severe lack of segregation
of class $X$ points from class $Y$ points (or class $Y$ points being significantly associated with class $X$ points),
and
$Z_{22}$ has smaller power since $Y$ points are clustered around $X$ points,
which also causes slight clustering of $Y$ points.

\textbf{Case III:}
For the  third class of alternatives, we consider
$X_i \stackrel{iid}{\sim} \U((0,1-s)\times(0,1-s))$ for $i=1,\ldots,n_1$,
and
$Y_j \stackrel{iid}{\sim} \U((s,1)\times(s,1))$ for $j=1,\ldots,n_2$.
The three values of $s$ constitute the following alternatives;
\begin{equation}
\label{eqn:Ha-case-III}
H_{III}^{1}: s=1/6,\;\;\;H_{III}^{2}: s=1/4, \text{ and } H_{III}^{3}: s=1/3.
\end{equation}
Notice that these alternatives are the segregation alternatives considered
for Monte Carlo analysis in \cite{ceyhan:overall}.
The empirical power estimates
under the segregation alternatives
are presented in Table \ref{tab:alternative-3}.
The exact tests have very low power.
The NN reflexivity and species-correspondence tests have high power
which increases as $s$ increases.
Furthermore,
$Z_{s,r}$ has high power for the right-sided alternative
and
$Z_{m,nr}$ has high power for the left-sided alternative,
which indicates significant self-reflexivity in NN structure.
Furthermore,
there is significant species-correspondence (at the same level for both classes by construction),
and the cell-specific tests are also significant for the right-sided alternatives
indicating significant segregation of the classes.

\begin{table}[ht]
\centering
\begin{tabular}{|c|c|c||c|c|c||c|c|c|}
\hline
\multicolumn{1}{|c}{ }&\multicolumn{8}{|c|}{Power estimates under the case III alternatives} \\
\hline
& $\bh_F^{>}$ & $\bh_F^{<}$ & $\bh_R$ &
$\bh^{Z,>}_{s,r}$ & $\bh^{Z,<}_{m,nr}$ & $\bh_{sc}$ & $\bh^{Z,>}_{11}$ & $\bh^{Z,>}_{22}$ \\
\hline
$H_{III}^1$ & .0745 & .0320 & .4554 & .4541 & .4233 & .4233 & .5179 & .5215\\
\hline
$H_{III}^2$ & .0613 & .0287 & .9451 & .8720 & .8709 & .9246 & .9460 & .9432\\
\hline
$H_{III}^3$ & .0524 & .0229 & .9999 & .9975 & .9952 & 1.000 & .9999 & .9998\\
\hline
\end{tabular}
\caption{
\label{tab:alternative-3}
The power estimates under the case III alternatives with $N_{mc}=10000$,
$n_1=n_2=40$ at $\alpha=.05$.
The empirical power labeling and superscripting for ``$<$" and ``$>$" are as in Table \ref{tab:alternative-1}.}
\end{table}

\textbf{Case IV:}
We also consider alternatives in which
self-reflexive pairs are more frequent than expected by construction.
We generate
$X_i\stackrel{iid}{\sim} S_1$ for $i=1,\ldots,\lfloor n_1/2 \rfloor$
and
$Y_j\stackrel{iid}{\sim} S_2$ for $j=1,\ldots,\lfloor n_2/2 \rfloor$.
Then for $k=\lfloor n_1/2 \rfloor+1,\ldots,n_1$,
we generate $X_k=X_{k-\lfloor n_1/2 \rfloor}+r\,(\cos T_j, \sin T_j)^t$
and for $l=\lfloor n_2/2 \rfloor+1,\ldots,n_2$,
we generate $Y_l=Y_{l-\lfloor n_1/2 \rfloor}+r\,(\cos T_j, \sin T_j)^t$
where $r \in (0,1)$ and $T_j \sim \U(0,2\,\pi)$.
Appropriate small choices of $r$ will
yield an abundance of self-reflexive pairs.
The three values of $r$ we consider
constitute the self-reflexivity alternatives at each support pair $(S_1,S_2)$.
Then the nine alternative combinations we consider are given by

\begin{align}
\label{eqn:Ha-case-IV}
(i) & \text{ $H_{IV}^1:$
$S_1=S_2=(0,1) \times (0,1)$, (a) $r=1/7$, (b) $r=1/8$, (c) $r=1/9$,} \nonumber \\
(ii) & \text{ $H_{IV}^2:$
$S_1=(0,5/6) \times (0,5/6)$
and $S_2=(1/6,1) \times (1/6,1)$,  (a) $r=1/7$, (b) $r=1/8$, (c) $r=1/9$,}\\
(iii) & \text{ $H_{IV}^3:$
$S_1=(0,3/4) \times (0,3/4)$
and $S_2=(1/4,1) \times (1/4,1)$  (a) $r=1/7$, (b) $r=1/8$, (c) $r=1/9$.} \nonumber
\end{align}

\begin{table}[ht]
\centering
\begin{tabular}{|c|c|c|c||c|c|c||c|c|c|}
\hline
\multicolumn{2}{|c}{ }&\multicolumn{8}{|c|}{Power estimates under the case IV alternatives} \\
\hline
 & $r$ & $\bh_F^{>}$ & $\bh_F^{<}$ & $\bh_R$ &
$\bh^{Z,>}_{s,r}$ & $\bh^{Z,<}_{m,nr}$ & $\bh_{sc}$ & $\bh^{Z,>}_{11}$ & $\bh^{Z,>}_{22}$ \\
\hline
 & 1/7 & .3340 & .0025 & .8713 & .9254 & .4444 & .8708 & .8894 & .8868\\
\cline{2-10}
$H_{IV}^1$ & 1/8 & .3903 & .0006 & .9391 & .9699 & .4955 & .9405 & .9451 & .9445\\
\cline{2-10}
 & 1/9 & .4377 & .0004 & .9726 & .9883 & .5276 & .9733 & .9741 & .9748\\
\hline
\hline
 & 1/7 & .2599 & .0042 & .9490 & .9640 & .6771 & .9478 & .9567 & .9576\\
\cline{2-10}
$H_{IV}^2$ & 1/8 & .3024 & .0019 & .9756 & .9871 & .7085 & .9767 & .9775 & .9792\\
\cline{2-10}
 & 1/9 & .3387 & .0013 & .9892 & .9943 & .7320 & .9882 & .9897 & .9910\\
\hline
\hline
 & 1/7 & .1965 & .0062 & .9914 & .9913 & .8883 & .9921 & .9918 & .9933\\
\cline{2-10}
$H_{IV}^3$ & 1/8 & .2289 & .0042 & .9961 & .9966 & .8924 & .9974 & .9955 & .9961\\
\cline{2-10}
 & 1/9 & .2640 & .0022 & .9984 & .9990 & .8959 & .9985 & .9974 & .9981\\
\hline
\end{tabular}
\caption{
\label{tab:alternative-4}
The power estimates under the case IV alternatives with $N_{mc}=10000$,
$n_1=n_2=40$ at $\alpha=.05$.
The empirical power labeling and superscripting for ``$<$" and ``$>$" are as in Table \ref{tab:alternative-1}.}
\end{table}

In this case,
under $H_{IV}^2$ and $H_{IV}^3$,
by construction,
there is species-correspondence and hence segregation of the classes due to the choices of the supports.
Additionally,
with decreasing $r$, the self-reflexive pairs will be more and more abundant.
The empirical power estimates under the self-reflexivity in NN structure alternatives
are presented in Table \ref{tab:alternative-4}.
Notice that
left-sided exact test on the NN-RCT
has almost no power.
The right-sided exact test on the NN-RCT has moderate power.
The NN reflexivity and species-correspondence tests all have very high power estimates.
Furthermore,
$Z_{s,r}$ has high power for the right-sided alternative
and
$Z_{m,nr}$ has high power for the left-sided alternative,
which indicates significant presence of self-reflexivity in NN structure.
Also,
there is significant species-correspondence (at the same level for both classes by construction),
and the cell-specific tests are also significant for the right-sided alternatives
indicating significant segregation of the classes.
The higher power estimates for these tests increase from $H_{IV}^1$ to $H_{IV}^3$
and also  they increase as $r$ decreases.
Hence the higher power estimates increase as the levels of species-correspondence and
self-reflexivity in NN structure increase.

\textbf{Case V:}
In this case, first,
we generate $X_i\stackrel{iid}{\sim} \U((0,1)\times(0,1))$
and then generate $Y_j$ as
$Y_j=X_i+r\,(\cos T_j, \sin T_j)^t$
where $r \in (0,1)$ and $T_j \sim \U(0,2\,\pi)$.
In the pattern generated,
appropriate choices of $r$ will cause
$Y_j$ and $X_i$ more associated,
that is,
a $Y$ point will be more likely to be the NN of an $X$ point,
and vice versa.
The three values of $r$ we consider
constitute the three association alternatives;
\begin{equation}
\label{eqn:Ha-case-V}
H_V^1: r=1/4,\;\;\; H_V^2: r=1/7, \text{ and } H_V^3: r=1/10.
\end{equation}
These are also the association alternatives considered
for Monte Carlo analysis in \cite{ceyhan:overall}.

\begin{table}[ht]
\centering
\begin{tabular}{|c|c|c||c|c|c||c|c|c|}
\hline
\multicolumn{1}{|c}{ }&\multicolumn{8}{|c|}{Power estimates under the case V alternatives} \\
\hline
& $\bh_F^{>}$ & $\bh_F^{<}$ & $\bh_R$ &
$\bh^{Z,<}_{s,r}$ & $\bh^{Z,>}_{m,nr}$ & $\bh_{sc}$ & $\bh^{Z,<}_{11}$ & $\bh^{Z,<}_{22}$ \\
\hline
$H_V^1$ & .0196 & .1597 & .1736 & .3222 & .1223 & .1897 & .2206 & .3909\\
\hline
$H_V^2$ & .0071 & .2631 & .4620 & .6423 & .2032 & .4541 & .5649 & .6347\\
\hline
$H_V^3$ & .0030 & .3799 & .7049 & .8443 & .2446 & .6808 & .7937 & .7795\\
\hline
\end{tabular}
\caption{
\label{tab:alternative-5}
The power estimates under the case V alternatives with $N_{mc}=10000$,
$n_1=n_2=40$ at $\alpha=.05$.
The empirical power labeling and superscripting for ``$<$" and ``$>$" are as in Table \ref{tab:alternative-1}.}
\end{table}

The empirical power estimates under $H_V^1-H_V^3$
are presented in Table \ref{tab:alternative-5}.
Notice that
right-sided exact test on the NN-RCT has virtually zero power.
The left-sided exact test on the NN-RCT has moderate power.
The $\chi^2$ tests of NN reflexivity and species-correspondence have high power
(which increases as $r$ decreases).
But
$Z_{s,r}$ has high power for the left-sided alternative
and
$Z_{m,nr}$ has high power for the right-sided alternative only,
which indicates significant lack of self-reflexivity in NN structure and presence of moderate mixed-nonreflexivity in the NN structure.
Also,
the cell-specific tests are also significant for the left-sided alternatives
indicating significant lack of segregation (or presence of significant association) of the classes.
The power estimates for these tests increase as $r$ decreases.

\section{Example Data: Urkiola Woods Data}
\label{sec:example-data}
\begin{figure}[ht]
\centering
%\psfrag{Density}{ \Huge{\bf{Density}}}
\rotatebox{-90}{ \resizebox{3.5 in}{!}{\includegraphics{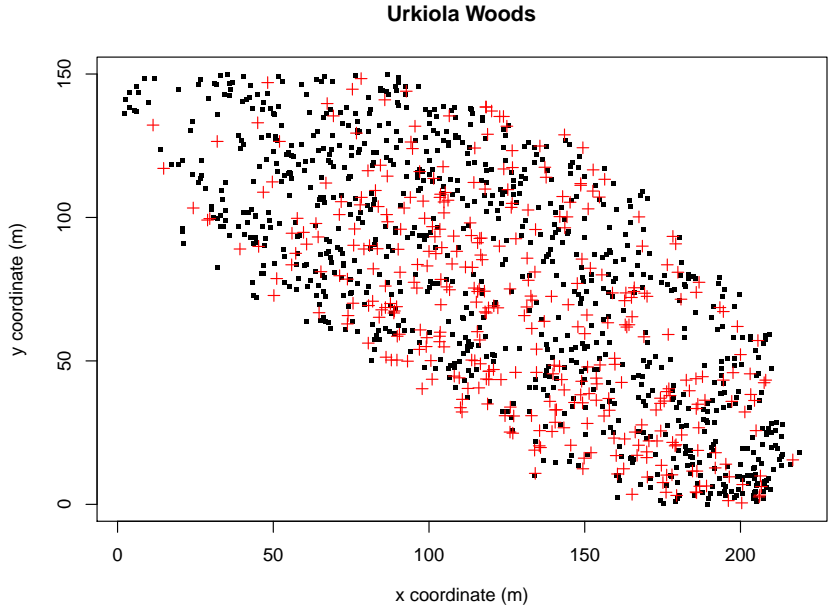} }}
 \caption{
\label{fig:UrkiolaWoods}
The scatter plot of the locations of birch trees (solid squares {\scriptsize  $\blacksquare$}),
and oak trees (pluses $+$) in the  Urkiola Natural Park, Basque region, northern Spain.}
\end{figure}

To illustrate the methods,
we use the Urkiola Woods data,
which contains locations of trees (in meters)
in a secondary wood in Urkiola Natural Park, Basque region, northern Spain (\cite{laskurain:Phd-thesis}).
The data set is available in the spatstat package in R (\cite{baddeley:2005}),
and contains 886 birch trees (Betula celtiberica) and 359 oak trees (Quercus robur).
This data set is actually a part of a more extensive data set collected and analyzed by \cite{laskurain:Phd-thesis}.
The scatter plot of the tree locations are presented in Figure \ref{fig:UrkiolaWoods}.

\begin{table}[h]
\centering
\begin{tabular}{cc|cc|c}
\multicolumn{2}{c}{}& \multicolumn{2}{c}{pair type}& \\
\multicolumn{2}{c}{} &  self pairs   & mixed pairs  &   total  \\
\hline
& reflexive pairs  &    475    &    259    &   734  \\
\raisebox{1.5ex}[0pt] {NN reflexivity}
& non-reflexive pairs &    323    &    188    &   511  \\
\hline
& total     &    798     &    447             &   1245 \\
\end{tabular}
\caption{
\label{tab:ref-CT-urkiola}
The NN-RCT for Urkiola Woods data.}
\end{table}

The NN-RCT for this data is presented in Table \ref{tab:ref-CT-urkiola}.
Notice that the ratio of the frequency of reflexive pairs to that of non-reflexive pairs is
$734/511 \approx 1.44$.
The same ratio among self pairs is $475/323 \approx 1.47$ and among mixed pairs is $259/188 \approx 1.38$,
which are very close to the overall ratio (of the row sums).
Hence,
there seems to exist independence between NN reflexivity and pair type as self or mixed
for the tree species in the Urkiola Woods data set.

\begin{table}[h]
\centering
\begin{tabular}{cc|cc|c}
\multicolumn{2}{c}{}& \multicolumn{2}{c}{pair type }& \\
\multicolumn{2}{c}{}&    self &  mixed   &   total  \\
\hline
& birch &    668    &    218    &   886  \\
\raisebox{1.5ex}[0pt] {base species}
&oak &    130    &  229    &   359  \\
\hline
&total     &    798   &   447       &  1245  \\
\end{tabular}
\caption{
\label{tab:NNCT-urkiola}
The SCCT for the Urkiola Woods data set containing birch and oak trees.}
\end{table}

The SCCT for this data set is presented in Table \ref{tab:NNCT-urkiola}.
The abundance ratio for the species is $886/359 \approx 2.47$
and the ratio of the entries in the self column is $ 668/130 \approx 5.14$,
which seems to be much larger than 2.47, suggesting presence of species-correspondence
(at least for one of the species).

\begin{table}[ht]
\centering
\begin{tabular}{|c|c|c|c||c|c||c|c|c||c|c|c|}
\hline
\multicolumn{1}{|c}{ }&\multicolumn{11}{|c|}{Test statistics and $p$-values for Urkiola Woods data} \\
\hline
& $\X^2_P$ & $Z^{>}_{dir}$ & $Z^{<}_{dir}$  & $T_F^{>}$ & $T_F^{<}$ & $\X^2_R$ & $Z^>_{sr}$ & $Z^<_{mn}$ & $\N_I$ & $Z^>_{11}$ & $Z^>_{22}$ \\
\hline
TS       & .2346 & \multicolumn{2}{|c||}{ .5444 } & \multicolumn{2}{|c||}{ 1.0674 } & 8.9538 & 2.2539 & -1.9682 & 11.4079 & 2.9011 & 2.7047 \\
\hline
$\pasy$  & .6282 & .2931 & .7069 & .3138* & .7274* & .0114 & .0121 & .0245 & .0033 & .0019 & .0034 \\
\hline
$\prand$ & .6251 & .3155 & .6845 & .3150 & .6850 & .0044 & .0070 & .0209 & .0032 & .0011 & .0043 \\
\hline
\end{tabular}
\caption{
\label{tab:p-val-urkiola}
The test statistics and the $p$-values for Urkiola Woods data.
$Z_{sr}$, $Z_{mn}$, $\X^2_P$, $\X^2_R$, $Z_{11}$, $Z_{22}$, and $\N_I$
are as defined in the text;
$Z^{>}_{dir}$ and $Z^{<}_{dir}$ are for the right-sided and left-sided directional test $Z_{dir}$;
$T_F^{>}$ and $T_F^{<}$ are one-sided Fisher's exact test (for the right and left-sided tests
on the NN-RCT, respectively).
TS stands for the test statistic,
$\pasy$ for the $p$-values based on asymptotic critical values (except for the exact tests)
and
$\prand$ for the $p$-values based on Monte Carlo randomization.
* The $p$-values for the exact tests are computed as described in Section \ref{sec:fisher-exact-test}.}
\end{table}

%The more appropriate null hypothesis is the CSR independence pattern,
%since the locations of the tree species can be viewed a priori resulting from different processes (\cite{goreaud:2003}).
We present the test statistics and the associated $p$-values in Table \ref{tab:p-val-urkiola},
where $Z_{sr}$, $Z_{mn}$, $\X^2_P$, $\X^2_R$, $Z_{11}$, $Z_{22}$, and $\N_I$
are as defined in the text,
and
$Z^{>}_{dir}$ and $Z^{<}_{dir}$ are for the right-sided and
left-sided versions of the directional test $Z_{dir}$.
Furthermore,
$T_F^{>}$ and $T_F^{<}$ are one-sided Fisher's exact test (for the right and left-sided tests
on the NN-RCT, respectively) where the test statistic is the odds ratio.
Furthermore,
in this table $\pasy$ stands for the $p$-value based on the asymptotic approximation (i.e., asymptotic critical value)
except for the exact tests,
$\prand$ is based on Monte Carlo
randomization of the labels on the given locations of the trees 10000 times.
For the exact tests, the $p$-value written for the $\pasy$ row is computed as in Section \ref{sec:fisher-exact-test}.
Notice that $\pasy$ and $\prand$ are similar for most tests.
For the tests with the correct or approximate asymptotic sampling distributions,
$\pasy$ and $\prand$ are very close (closest for the cell-specific and species-correspondence tests).

Notice that
Pearson's $\chi^2$ test of independence and the corresponding one-sided exact tests on the
NN-RCT suggest no significant deviation from independence.
However, among these tests,
the asymptotic ones do not have the correct sampling distribution,
and the exact tests are valid for small sample sizes (less than about 50).
Hence the asymptotic approximation and the exact tests would not be reliable.
The Monte Carlo randomized $p$-value, $\prand$, is very similar to $\pasy$ values,
suggesting independence between NN reflexivity and pair type as self or mixed.
On the other hand,
the $Z$-test for self-reflexivity in NN structure is significant for the right-sided alternative
and
mixed-nonreflexivity is significant for the left-sided alternative
and
$\chi^2$-test for NN reflexivity, $\X^2_R$, are all significant,
implying presence of strong self-reflexivity in NN structure.
Likewise,
the cell-specific tests are both significant for the right-sided alternative,
and
the $\chi^2$ species-correspondence test, $\N_I$, is significant,
implying significant species-correspondence for these species,
and hence significant segregation of the species (from each other).

\section{Discussion and Conclusions}
\label{sec:disc-and-conc}
In this article,
we discuss various tests of spatial interaction in the NN structure based on contingency tables.
In particular,
we investigate tests of NN reflexivity and species-correspondence
using contingency tables based on the NN relations between classes or species.
We consider Pielou's test proposed for niche specificity (\cite{pielou:1961}),
determine its appropriate null hypothesis and the underlying assumptions
and demonstrate that Pielou's contingency table intended for niche specificity is
actually more appropriate for NN reflexivity
(hence called as NN reflexivity contingency table (NN-RCT) in this article).
As an alternative,
we provide an approximate asymptotic distribution to the entries of the NN-RCT
and thus propose new tests of NN reflexivity.
Pearson's $\chi^2$ test of independence (suggested by \cite{pielou:1961})
and the one-sided versions
on the RCT are slightly liberal with the asymptotic approximation,
but our new NN reflexivity tests are about the desired level.
We also introduce a new test of species-correspondence
and the associated contingency table called species-correspondence contingency table (SCCT)
which is derived from NNCT for this purpose.
Self-reflexivity in NN structure can account for segregation as can species-correspondence
and niche specificity.
In the presence of segregation,
if the supports of the classes are about the same,
then self-reflexivity in NN structure accounts more for segregation.
If the supports of the classes are considerably different,
niche specificity or species-correspondence accounts for segregation,
but still
self-reflexivity in NN structure might partially account for segregation.
We also consider the use of Fisher's exact test on the contingency tables
and based on our extensive Monte Carlo simulations,
although one version of the exact tests has the appropriate empirical level,
exact tests are not the best performers in terms of power,
hence are not recommended for use in practice for NN-RCT.
In particular,
we demonstrate that table inclusive versions of the one-sided tests
are more appropriate for the NN-RCT
when class sizes are small (i.e., less than about 40).

In the literature
usually NN relationships are based on the distance metrics.
For example, in this article,
Euclidean distance in $\R^2$ is the only metric used.
The NN relations based on dissimilarity measures is an extension of NN relations based on distance metrics.
In such an extension, NN of an object, $x$, refers to the object
with the minimum dissimilarity to $x$.
We assume that the objects (events) lie in a finite or infinite dimensional space
satisfying the lack of any inter-dependence which imply self- or mixed-reflexivity in NN structure.
Under RL,
the objects are fixed yielding fixed interpoint
dissimilarity measures, but the labels are assigned randomly to the objects.
The extensions of Pielou's test of independence and
our newly proposed test on the NN-RCT are straightforward.
However our species-correspondence tests are constructed assuming data are in $\R^2$.
In particular, the quantity $Q$ which is the number of points with
shared NNs needs to be updated for higher dimensional data.
The form of $Q$ in $\R^2$ is defined as
$\widetilde Q:=2\,\sum_{j=1}^n {j \choose 2} Q_j$.
Usually we have $\widetilde Q \approx Q$ in practice.
One may check the validity of this assumption
by using the interpoint dissimilarity matrix
in the classical multi-dimensional scaling (\cite{cox:2001}) of the data to $\R^2$.
If the NN relations remain similar,
it might be more practical to just use $Q$ instead of $\widetilde Q$
for computational purposes.
Furthermore,
a point can serve as a NN to more than 6 points
with non-Euclidean distances or dissimilarity measures.

\section*{Acknowledgments}
%I would like to thank an anonymous associate editor and two referees,
%whose constructive comments and suggestions greatly improved the presentation
%and flow of the paper.
%Most of the Monte Carlo simulations presented in this article
%were executed at Ko\c{c} University High Performance Computing Laboratory.
This research was supported by the research agency TUBITAK via Project \# 111T767
and
by the European Commission under the
Marie Curie International Outgoing Fellowship Programme
via Project \# 329370 titled PRinHDD.

%\bibliography{References}

\begin{thebibliography}{}

\bibitem[Agresti, 1992]{agresti:1992}
Agresti, A. (1992).
\newblock A survey of exact inference for contingency tables.
\newblock {\em Statistical Science}, 7:131--153.

\bibitem[Baddeley and Turner, 2005]{baddeley:2005}
Baddeley, A.~J. and Turner, R. (2005).
\newblock spatstat: An {R} package for analyzing spatial point patterns.
\newblock {\em Journal of Statistical Software}, 12(6):1--42.

\bibitem[Benayas et~al., 1999]{benayas:1999}
Benayas, R. J.~M., Scheiner, S.~M., S\'{a}nchez-Colomer, M.~G., and Levassor,
  C. (1999).
\newblock Commonness and rarity: theory and application of a new model to
  {M}editerranean montane grasslands.
\newblock Conservation Ecology 3(1): 5. Available online at URL:
  \url{http://www.consecol.org/vol3/iss1/art5}.

\bibitem[Ceyhan, 2008]{ceyhan:cell2008}
Ceyhan, E. (2008).
\newblock Overall and pairwise segregation tests based on nearest neighbor
  contingency tables.
\newblock {\em Computational Statistics \& Data Analysis}, 53(8):2786--2808.

\bibitem[Ceyhan, 2010a]{ceyhan:exact-NNCT}
Ceyhan, E. (2010a).
\newblock Exact inference for testing spatial patterns by nearest neighbor
  contingency tables.
\newblock {\em Journal of Probability and Statistical Science}, 8(1):45--68.

\bibitem[Ceyhan, 2010b]{ceyhan:overall}
Ceyhan, E. (2010b).
\newblock On the use of nearest neighbor contingency tables for testing spatial
  segregation.
\newblock {\em Environmental and Ecological Statistics}, 17(3):247--282.

\bibitem[Chuyong et~al., 2011]{chuyong:2011}
Chuyong, G.~B., Kenfack, D., Harms, K.~E., Thomas, D.~W., Condit, R., and
  Comita, L.~S. (2011).
\newblock Habitat specificity and diversity of tree species in an {A}frican wet
  tropical forest.
\newblock {\em Plant Ecology}, 212:1363--1374.

\bibitem[Cox and Hinkley, 1974]{cox:1974}
Cox, D.~R. and Hinkley, D.~V. (1974).
\newblock {\em Theoretical Statistics}.
\newblock Chapman \& Hall / CRC, Boca Raton, FL.

\bibitem[Cox and Cox, 2001]{cox:2001}
Cox, T. and Cox, M. (2001).
\newblock {\em Multidimensional Scaling}.
\newblock Chapman and Hall, Boca Raton, FL.

\bibitem[Diggle, 1979]{diggle:1979}
Diggle, P.~J. (1979).
\newblock On parameter-estimation and goodness-of-fit testing for spatial point
  patterns.
\newblock {\em Biometrics}, 35(1):87--101.

\bibitem[Dixon, 1994]{dixon:1994}
Dixon, P.~M. (1994).
\newblock Testing spatial segregation using a nearest-neighbor contingency
  table.
\newblock {\em Ecology}, 75(7):1940--1948.

\bibitem[Dixon, 2002a]{dixon:NNCTEco2002}
Dixon, P.~M. (2002a).
\newblock Nearest-neighbor contingency table analysis of spatial segregation
  for several species.
\newblock {\em Ecoscience}, 9(2):142--151.

\bibitem[Dixon, 2002b]{dixon:EncycEnv2002}
Dixon, P.~M. (2002b).
\newblock Nearest neighbor methods.
\newblock {\em Encyclopedia of Environmetrics, edited by Abdel H. El-Shaarawi
  and Walter W. Piegorsch, John Wiley \& Sons Ltd., NY}, 3:1370--1383.

\bibitem[Freeman, 2002]{freeman:2002}
Freeman, S. (2002).
\newblock {\em Biological Science}.
\newblock Prentice Hall, Upper Saddle River, NJ.

\bibitem[Harms et~al., 2001]{harms:2001}
Harms, K.~E., Condit, R., Hubbell, S.~P., and Foster, R.~B. (2001).
\newblock Habitat associations of trees and shrubs in a 50-ha neotropical
  forest plot.
\newblock {\em Journal of Ecology}, 89:947--959.

\bibitem[Hosack et~al., 2006]{hosack:2006}
Hosack, G.~R., Dumbauld, B.~R., Ruesink, J.~L., and Armstrong, D.~A. (2006).
\newblock Habitat associations of estuarine species: Comparisons of intertidal
  mudflat, seagrass ({Z}ostera marina), and oyster ({C}rassostrea gigas)
  habitats.
\newblock {\em Estuaries and Coasts}, 26(6B):1150--1160.

\bibitem[Jiangshan et~al., 2009]{jiangshan:2009}
Jiangshan, L., Xiangcheng, M., Haibao, R., and Keping, M. (2009).
\newblock Species-habitat associations change in a subtropical forest of
  {C}hina.
\newblock {\em Journal of Vegetation Science}, 20:415--423.

\bibitem[Kulldorff, 2006]{kulldorff:2006}
Kulldorff, M. (2006).
\newblock Tests for spatial randomness adjusted for an inhomogeneity: A general
  framework.
\newblock {\em Journal of the American Statistical Association},
  101(475):1289--1305.

\bibitem[Laskurain, 2008]{laskurain:Phd-thesis}
Laskurain, N.~A. (2008).
\newblock {\em Din\'{a}mica espacio-temporal de un bosque secundario en el
  Parque Natural de Urkiola (Bizkaia).}
\newblock PhD thesis, Universidad del Pa\'{\i}s Vasco/Euskal Herriko
  Unibertsitatea.

\bibitem[Lindenmayer and Burgman, 2005]{lindenmayer:2005}
Lindenmayer, D. and Burgman, M. (2005).
\newblock {\em Practical Conservation Biology}.
\newblock {CSIRO} Publishing, Collingwood, VIC, Australia.

\bibitem[Pei et~al., 2011]{pei:2011}
Pei, N., Lian, J.-Y., Erickson, D.~L., Swenson, N.~G., Kress, W.~J., Ye, W.-H.,
  and Ge, X.-J. (2011).
\newblock Exploring tree-habitat associations in a {C}hinese subtropical forest
  plot using a molecular phylogeny generated from {DNA} barcode loci.
\newblock {\em {PLoS ONE}}, 6(6):e21273. doi:10.1371/journal.pone.0021273.

\bibitem[Pielou, 1961]{pielou:1961}
Pielou, E.~C. (1961).
\newblock Segregation and symmetry in two-species populations as studied by
  nearest-neighbor relationships.
\newblock {\em Journal of Ecology}, 49(2):255--269.

\bibitem[Ramsey and Sjamsoe'oed, 1994]{ramsey:1994}
Ramsey, F.~L. and Sjamsoe'oed, R. (1994).
\newblock Habitat association studies in conjunction with adaptive cluster
  samples.
\newblock {\em Environmental and Ecological Statistics}, 1(2):121--132.

\bibitem[Ranker and Haufler, 2008]{ranker:2008}
Ranker, T.~A. and Haufler, C. H. (Editors). (2008).
\newblock {\em Biology and Evolution of Ferns and Lycophytes}.
\newblock Cambridge University Press, Cambridge, UK.

\bibitem[Ripley, 2004]{ripley:2004}
Ripley, B.~D. (2004).
\newblock {\em Spatial Statistics, 2nd edition}.
\newblock Wiley-{I}nterscience, New York.

\bibitem[Searle, 2006]{searle:2006}
Searle, S.~R. (2006).
\newblock {\em Matrix Algebra Useful for Statistics}.
\newblock Wiley-{I}nterscience, New York.

\bibitem[Tocher, 1950]{tocher:1950}
Tocher, K.~D. (1950).
\newblock Extension of the {N}eyman-{P}earson theory of tests to discontinuous
  variates.
\newblock {\em Biometrika}, 37:130--144.

\bibitem[van Lieshout and Baddeley, 1999]{lieshout:1999}
van Lieshout, M.~N.~M. and Baddeley, A.~J. (1999).
\newblock Indices of dependence between types in multivariate point patterns.
\newblock {\em Scandinavian Journal of Statistics}, 26:511--532.

\end{thebibliography}
%\bibliographystyle{apalike}
%\bibliographystyle{plain}

\end{document}